\documentclass[usenatbib]{mn2e}
\usepackage{graphicx}
\usepackage[usenames]{color}
\usepackage{times}
\def\gtsim{\lower.5ex\hbox{$\; \buildrel > \over \sim \;$}}

\title[GAMA: Green valley galaxy timescales]{Galaxy And Mass Assembly (GAMA): Timescales for galaxies crossing the green valley}
\author[S. Phillipps et al.]
{S. Phillipps$^1$, M.N. Bremer$^1$, A.M. Hopkins$^2$,  R. De Propris$^3$, E.N. Taylor$^4$, P.A. James$^5$, 
\and L.J.M. Davies$^6$, M.E. Cluver$^{4,7}$, S. P. Driver$^6$, S.A. Eales$^8$, B.W. Holwerda$^9$, 
\and L.S. Kelvin$^{5,10}$ and A.E. Sansom$^{11}$ 
\\
$^1$Astrophysics Group, School of Physics, University of Bristol, Tyndall Avenue, Bristol, BS8 1TL, UK\\
$^2$Australian Astronomical Optics, Macquarie University, 105 Delhi Road, North Ryde, NSW2113, Australia\\
$^3$ Finnish Centre for Astronomy with ESO, University of Turku, Vaisalantie 20, Piikkio, Finland\\
$^4$Centre for Astrophysics and Supercomputing, Swinburne University of Technology, Hawthorn, VIC3122, Australia\\
$^5$ Astrophysics Research Institute, Liverpool John Moores University, 146 Brownlow Hill, Liverpool L3 5RF, UK\\
$^6$ International Centre for Radio Astronomy Research, University of Western Australia, M468, 35 Stirling Highway, Crawley, WA 6009, Australia\\
$^7$ Department of Physics and Astronomy, University of Western Cape, Bellville 7535, South Africa\\
$^8$ School of Physics and Astronomy, Cardiff University, The Parade, Cardiff CF24 3AA, UK\\
$^9$ Department of Physics and Astronomy, University of Louisville, Louisville KY 40292, USA\\
$^{10}$ Department of Physics, University of California, One Shields Ave., Davis, CA 95616, USA\\
$^{11}$ Jeremiah Horrocks Institute, School of Physical Sciences and Computing, University of Central Lancashire, Preston PR1 2HE, UK}

\begin{document}

\date{Accepted . Received ; in original form }

\pagerange{\pageref{firstpage}--\pageref{lastpage}} \pubyear{}

\maketitle

\label{firstpage}

\begin{abstract}
We explore the constraints that can be placed on the evolutionary timescales for typical low redshift galaxies evolving from the blue cloud through the green valley and onto the red sequence. We utilise galaxies from the GAMA survey with $0.1<z<0.2$ and classify them according to the intrinsic ($u^*-r^*$) colours of their stellar populations, as determined by fits to their multi-wavelength spectral energy distributions. Using these fits to also determine stellar population ages and star formation timescales, we argue that our results are consistent with a green valley population dominated by galaxies that are simply decreasing their star formation (running out of gas) over a timescale of 2-4~Gyr which are seen at a specific epoch in their evolution (approximately 1.6 e-folding times after their peak in star formation). If their fitted star formation histories are extrapolated forwards, the green galaxies will further redden over time, until they attain the colours of a  passive population. In this picture, no specific quenching event which cuts-off their star formation is required, though it remains possible that the decline in star formation in green galaxies may be expedited by internal or external forces. However, there is no evidence that green galaxies have recently changed their star formation timescales relative to their previous longer term star formation histories.

\end{abstract}


\section{Introduction}

A feature of low redshift galaxies (in particular) is the dichotomy between `red' and `blue' galaxy populations, established in its modern context by \cite{Strateva2001}; see also \cite{Baldry2004}, \cite{Driver2006} and many others. Between the well populated `red sequence' and `blue cloud' populations in optical or ultraviolet colour-magnitude diagrams is the sparser `green valley' population, first discussed by \cite{Martin2007} and \cite{Wyder2007}. As colour must, at least roughly, reflect the specific star formation rate (sSFR), the star formation per unit existing stellar mass, these green galaxies have subsequently been studied primarily as likely examples of systems in transition from actively star forming to passively evolving, as in e.g. \cite{Fang2013} and the review by \cite{Salim2014}. 

Further, \cite{Bell2003} observed that at $z<1$ stars form predominantly in blue galaxies, but stellar mass accumulates in red galaxies, a point recently emphasised by \cite{Eales2015}. This demonstrates the importance of galaxies transitioning through the green valley, from blue to red, to our understanding of the last $\sim$~7~Gyr of galaxy evolution in general.

Indeed, while it is possible that some green galaxies have become green by the rejuvenation of star formation in a previously passive galaxy \citep[e.g.][]{Thilker2010, Mapelli2015}, it is apparent that as star formation fades, or is quenched, galaxies with intermediate levels of star formation must in general have green colours at some point while transitioning towards being `red and dead' \citep[][henceforth Paper I, and references therein]{Bremer2018}.

Following from this, key questions are then whether the green valley represents a specific phase of the evolution of typical galaxies, and if so, how long does this phase last?

To try to answer these questions and explore potential transition populations in general, as in Paper I we here take advantage of the rich panchromatic Galaxy And Mass Assembly (GAMA) survey data for a highly spectroscopically complete set of low redshift galaxies \citep{Driver2011, Liske2015}. We exploit, in particular, fits to the multi-wavelength spectral energy distributions (SEDs) made using stellar population synthesis techniques, which provide star formation rates (SFRs) and parameterised star formation histories (SFHs). This information allows us to investigate, for example, the evolutionary stage reached by green valley galaxies, how their evolution has proceeded in the recent past, and how it is likely to proceed into the future.

All magnitudes used in this work are in the AB system. Where relevant we use H$_0$ = 70~kms$^{-1}$Mpc$^{-1}$, $\Omega_{\rm m} = 0.3$ and $\Omega_\Lambda = 0.7$ as in \cite{Taylor2011}, from whose (updated) catalogue we take the GAMA masses, rest frame magnitudes, dust-corrected colours and fitted stellar population parameters.

\section{Sample selection}
\label{samp}
The GAMA survey provides a multi-wavelength database \citep{Driver2011, Driver2016} based on a highly complete galaxy redshift survey \citep{Baldry2010, Hopkins2013, Liske2015} covering approximately 280 deg$^2$ to a main survey magnitude limit of $r < 19.8$. This area is split into three equatorial (G09, G12 and G15) and two southern (G02 and G23) regions. The spectroscopic survey was undertaken with the AAOmega fibre-fed spectrograph \citep{Saunders2004, Sharp2006} allied to the Two-degree Field (2dF) fibre positioner on the Anglo-Australian Telescope \citep{Lewis2002}. It obtained redshifts for $\sim 250,000$ targets covering $0<z <0.6$ with a median redshift of $z\simeq 0.2$ with high (and uniform) spatial completeness on the sky in all the GAMA areas \citep{Robotham2010,Liske2015}. 

Built around the redshift survey, photometric data are provided at a wide range of wavelengths from the far-ultraviolet to far-infrared. Full details can be found in \cite{Driver2011, Driver2016}, \cite{Liske2015} and \cite{Baldry2018}. In this work we utilise the full GAMA II sample from the equatorial regions. 

Within the multi-wavelength database, each galaxy is characterised by a wide range of  observed and derived parameters, both photometric and spectroscopic. Those of interest in the current work include stellar masses, intrinsic extinction-corrected rest-frame colours and other stellar  population parameters, derived from multi-wavelength SED fitting \citep{Taylor2011}, as well as SFRs derived from various spectral line analyses or photometric measures \citep{Hopkins2013, Davies2016b, Wright2016}.

Following the method in \cite{Taylor2011}, the updated GAMA multi-band SED fitting is carried out using data available in the optical (SDSS) $ugriz$ and the VISTA VIKING near-IR $ZYJHK$ \citep{Edge2013} bands. The VIKING observations replaced the UKIDSS near-IR data discussed, but not ultimately used, in \cite{Taylor2011}. The fits themselves are made across the {\it rest-frame} wavelength range 3000-10000 \AA $\;$ and include allowance for the effects of dust extinction via a \cite{Calzetti2000} extinction curve. A total of 195669 galaxies have high quality spectra ($nQ>2$ in the GAMA catalogues) and fitted stellar masses (in the catalogue StellarMassesv19). 

The current investigation uses a sample of GAMA galaxies tuned for our specific priorities. The primary sample is seelected from the three equatorial GAMA regions and limited to the redshift range $0.1<z<0.2$. It is essentially volume limited for moderate to high luminosity galaxies ($M_r < -20$), and samples down to $M_r = -18.5$ at the low $z$ end. This overall sample contains 62188 galaxies. As in Paper I, the limited (fairly low) redshift range means that there is only a small differential in look-back time ($\sim 1$~Gyr), and hence potential evolution, {\it within} the sample.

To be conservative, we again follow Paper I and further restrict our samples to include only systems with $r$-band derived axial ratios $b/a>0.5$, in order to exclude galaxies with edge-on disks. These potentially have less reliable intrinsic photometry, given the large and possibly more uncertain dust corrections which are made necessary by strong inclination dependent obscuration \citep[e.g.][]{Driver2007}. The adopted `screen' model of the dust extinction as used in \cite{Taylor2011}, and most other similar work, will also be less appropriate for stars in edge-on disks \citep{Disney1989, Baes2001}. This leaves 39154 `round' galaxies between $z=0.1$ and 0.2.

In addition, as a check on any mass dependent effects, we also select a mass limited subsample of these galaxies which have stellar masses in the range  $10.25 <$ log($M_*/M_{\odot}$) $< 10.75$, the same range as used in Paper I. Finally, in Section 3.5, we briefly compare to a more local sample of 805 objects at $0.02 < z < 0.06$, otherwise limited in similar ways to our main sample \citep[see][henceforth Paper II]{Kelvin2018}.

\subsection{Galaxy Colours}
As in Paper I, the main sample is then split into three colour-selected subsamples -- `red', `green' and `blue' -- in terms of their intrinsic stellar, i.e. dust corrected, rest-frame, $(u^*-r^*)$ colours. Specifically, we divide our sample into these three broad colour bins based on the surface density of points in the intrinsic ($u^*-r^*$) colour-mass plane, utilising the colours of the best-fitting SED \citep{Taylor2011}. 

We should note that, while our chosen colour is clearly sensitive to the true evolutionary state of the stellar population, as required, the specification of a `green valley' in general terms is not clear cut.  The original definition came from GALEX observations in the ultraviolet, as discussed by \cite{Wyder2007}, \cite{Martin2007} and \cite{Salim2007}, amongst others, and selecting in terms of different optical or ultraviolet colours will likely produce different `green' samples \citep[see][]{Salim2014}.

Nevertheless, intrinsic $(u^*-r^*)$ colours do allow us to identify objects with stellar populations intermediate between purely passive and those undergoing so-called ``main sequence" star formation \citep{Noeske2007, Speagle2014} which dominate in the blue cloud. Evidently, any galaxy transitioning away from the main sequence of star formation must pass through this green colour regime at some point. In addition, contamination by dusty star forming galaxies, which could appear in samples defined only by {\em observed} colours  \citep[e.g.][]{SanchezAlmeida2010, Casado2015} should be minimised.

As in Paper I, we classify galaxies as red if they are on the red side of a straight line passing through ($u^*-r^*) = 1.5$ at log($M_*/M_{\odot}$) = 10 and  ($u^*-r^*) = 1.7$ at log($M_*/M_{\odot}$) = 11. This selects all the objects seen in the red sequence over this mass range and we merely extend it to higher and lower masses.  Blue galaxies are defined as those bluer than a line connecting ($u^*-r^*) =1.3$ to 1.4 over the same mass range, again covering the parameter space where sources typically classified as blue cloud galaxies reside. Galaxies are denoted as green between these two lines. The midpoint of these two lines essentially  follows the minimum surface density of objects in colour as a function of mass, i.e. the centre of the `green valley' as normally defined for these bands.

\begin{figure}
\includegraphics[width=\linewidth]{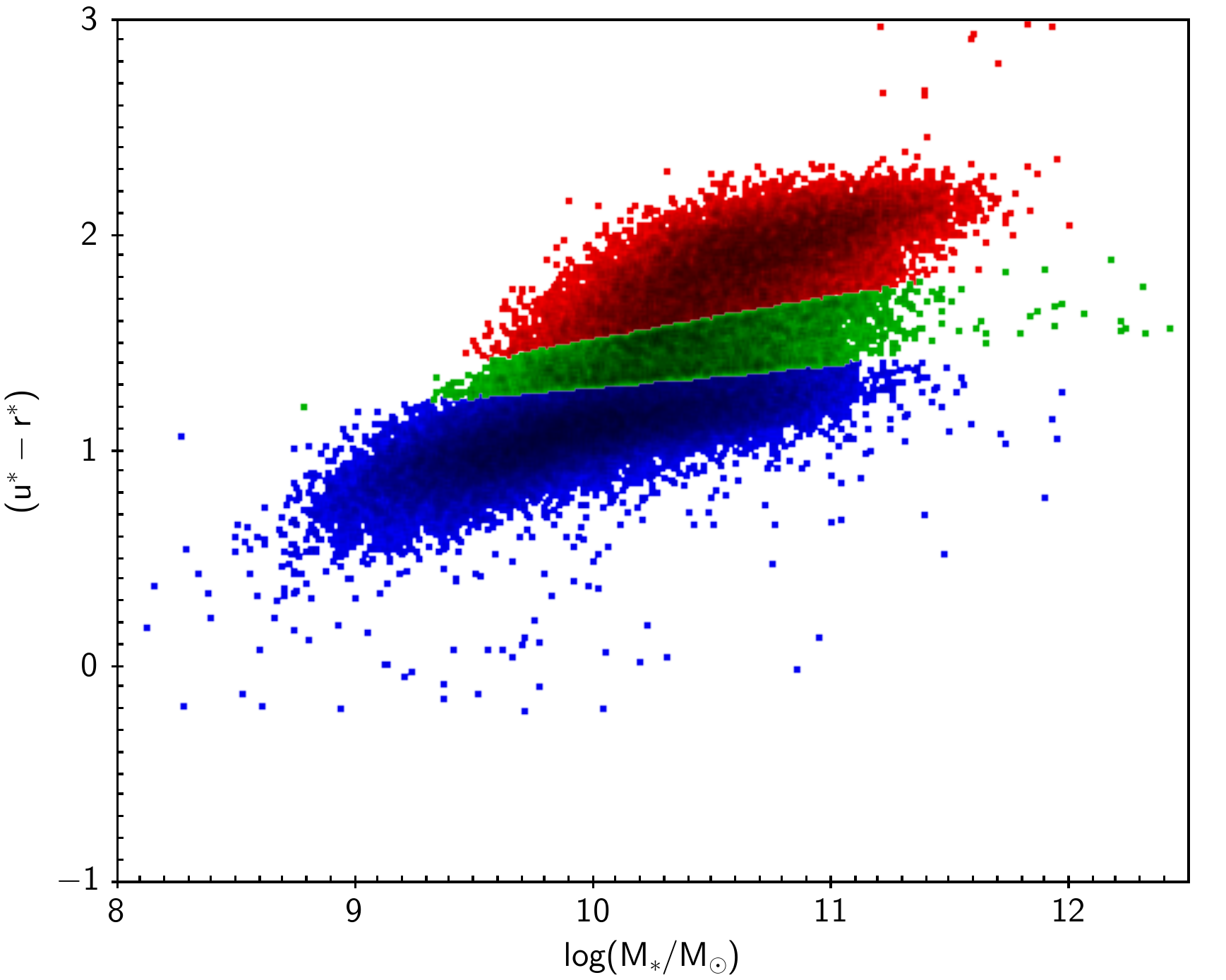}
\caption{The stellar mass versus intrinsic stellar population ($u^*-r^*$) colour plane for all non-edge-on (`round'), $0.1<z<0.2$ GAMA galaxies. The division into `red', `green' and 'blue' subsets (plotted in the appropriate colours) is as in Paper I, but extended to a wider mass range (see text).}
\label{time_mass_colour}
\end{figure}

\begin{figure}
\includegraphics[width=\linewidth]{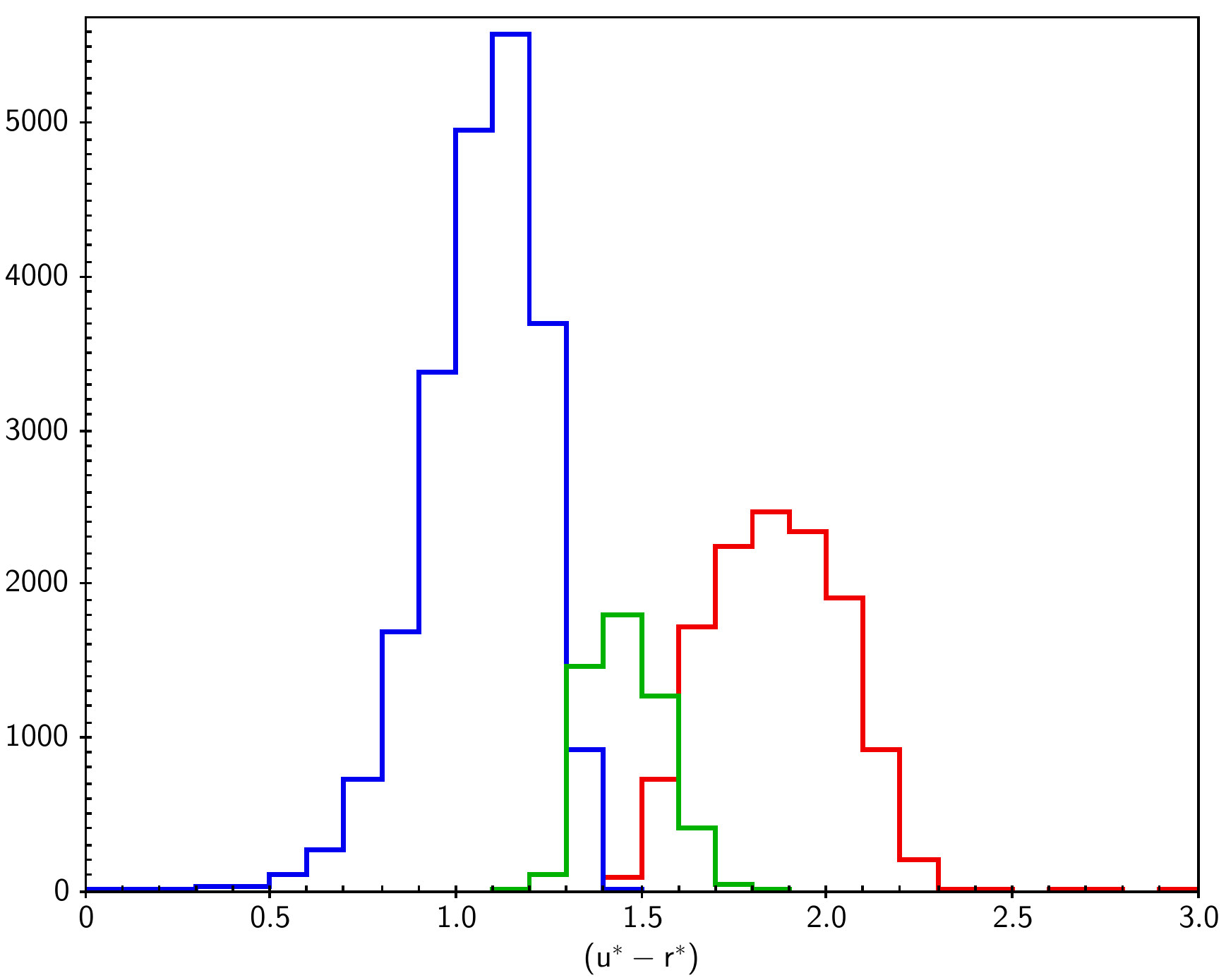}
\caption{Histograms of the ($u^* - r^*$) colours for each colour subsample. The subsamples overlap slightly because of the slope of the green valley in the colour-mass plane.}
\label{time_colour_hist}
\end{figure}

Our overall colour-mass plane is shown in Fig. \ref{time_mass_colour}. Our blue, green and red subsets contain 21406, 5068 and 12628 galaxies, respectively (ignoring a handful of objects outside the plotted limits). Histograms of the colour distributions for the red, green and blue galaxies are shown in Fig. \ref{time_colour_hist}. There is a small amount of overlap because of the mass dependence of the chosen edges of the green valley.

Note that the mass distribution, spanning $10^8$ to $10^{12.5} M_{\odot}$, varies between subsamples exactly as expected, with the blue fraction increasing towards lower masses and the red fraction towards higher masses \citep[e.g.][et seq.]{Kauffmann2003a, Kauffmann2003b, Baldry2004}. We have checked that this is not an artefact (circularity) of the method for determining $M_*$ based on the SED (and therefore colour), by alternatively plotting colour against $K-$band luminosity \citep[a good proxy for mass;][]{Bell2003}, obtaining entirely equivalent results. The wide mass range accounts for the lower green galaxy fraction here (13\%) than in Paper I ($\simeq$ 20 \%), which concentrated on a narrow intermediate mass range.

While our colour cuts can be considered somewhat arbitrary -- as noted above there is no single definition as to what constitutes a green valley galaxy -- as in Paper I we will also consider galaxy properties across the continuum of intrinsic colours and use the three broad colour bins for primarily descriptive purposes. This is similar to the approach of \cite{Oemler2017} who divide their galaxies into subsamples via their sSFR values.

In fact, the $(u^* - r^*)$ intrinsic colour is known to correlate reasonably well, but not perfectly, with the sSFR (see, e.g., Fig. 2 from Paper I). The GAMA database provides multiple derivations of the SFR \citep[see the summary in][]{Davies2016b}. In the present paper (again following Paper I), we utilise the SFRs obtained by use of the MAGPHYS code \citep{daCunha2008} as adapted for use with GAMA \cite[see][]{Driver2016, Wright2016}. Further, there are measures of the SFH of each galaxy. These are again obtained from the multi-band SED via a simple parameterized fit \citep{Taylor2011}, as discussed below.
An alternative, and complementary, route towards the stellar population parameters, and specifically the SFH, would be via fitting models to the galaxy spectra \citep[e.g.][and references therein]{Tojeiro2007,Wilkinson2017}, but this is beyond the scope of the present paper. In any case, as we are interested in galaxies subdivided by their broadband colours, it is more direct for our purposes to use models fitted to their SEDs.

\begin{figure}
\includegraphics[width=\linewidth]{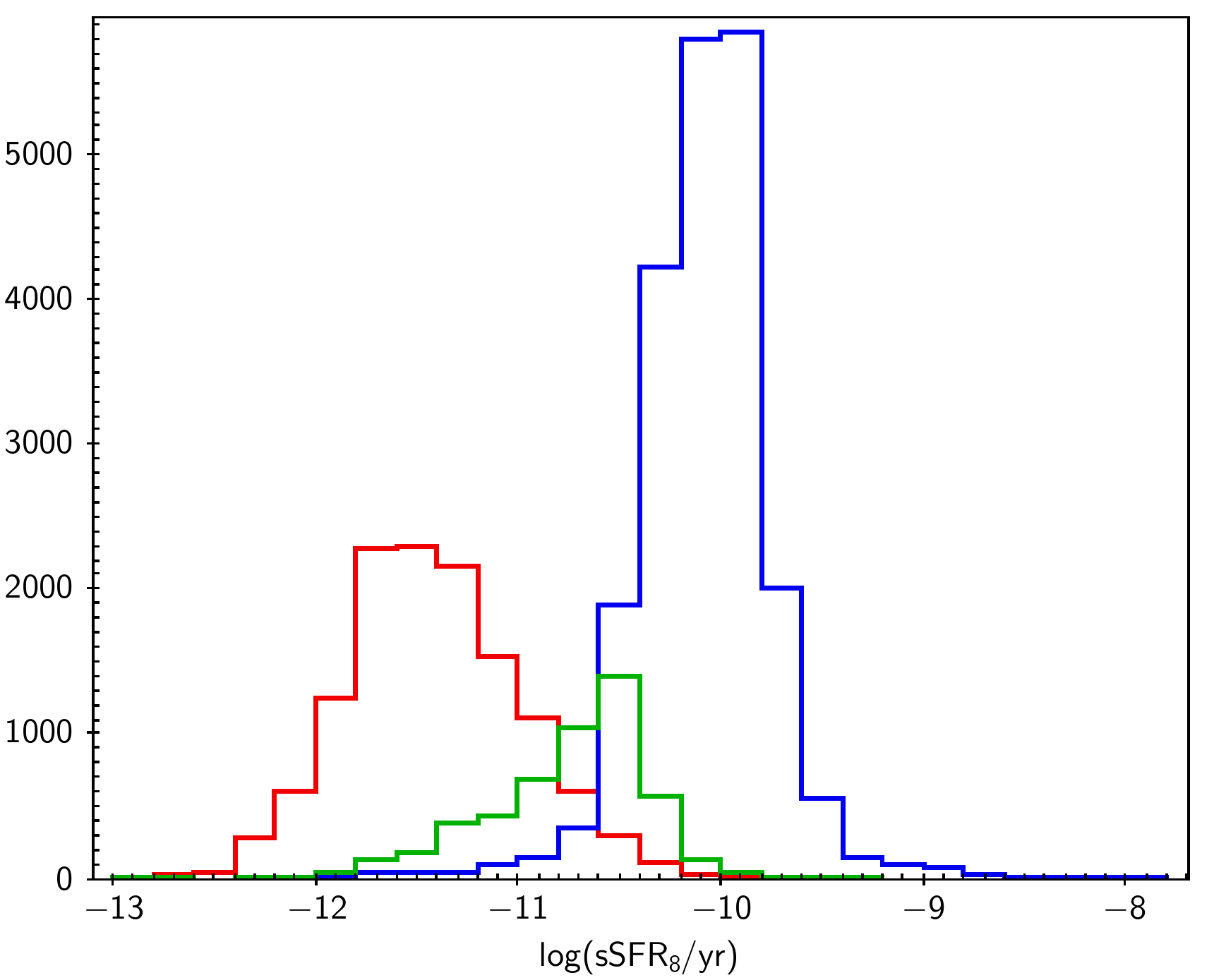}
\caption{Histogram of specific star formation rate for the three colour subsamples. Here, the star formation is the value averaged over the last $10^8$ years as derived via MAGPHYS (see Driver et al. 2016).}
\label{time_ssfr}
\end{figure}

\begin{figure}
\includegraphics[width=\linewidth]{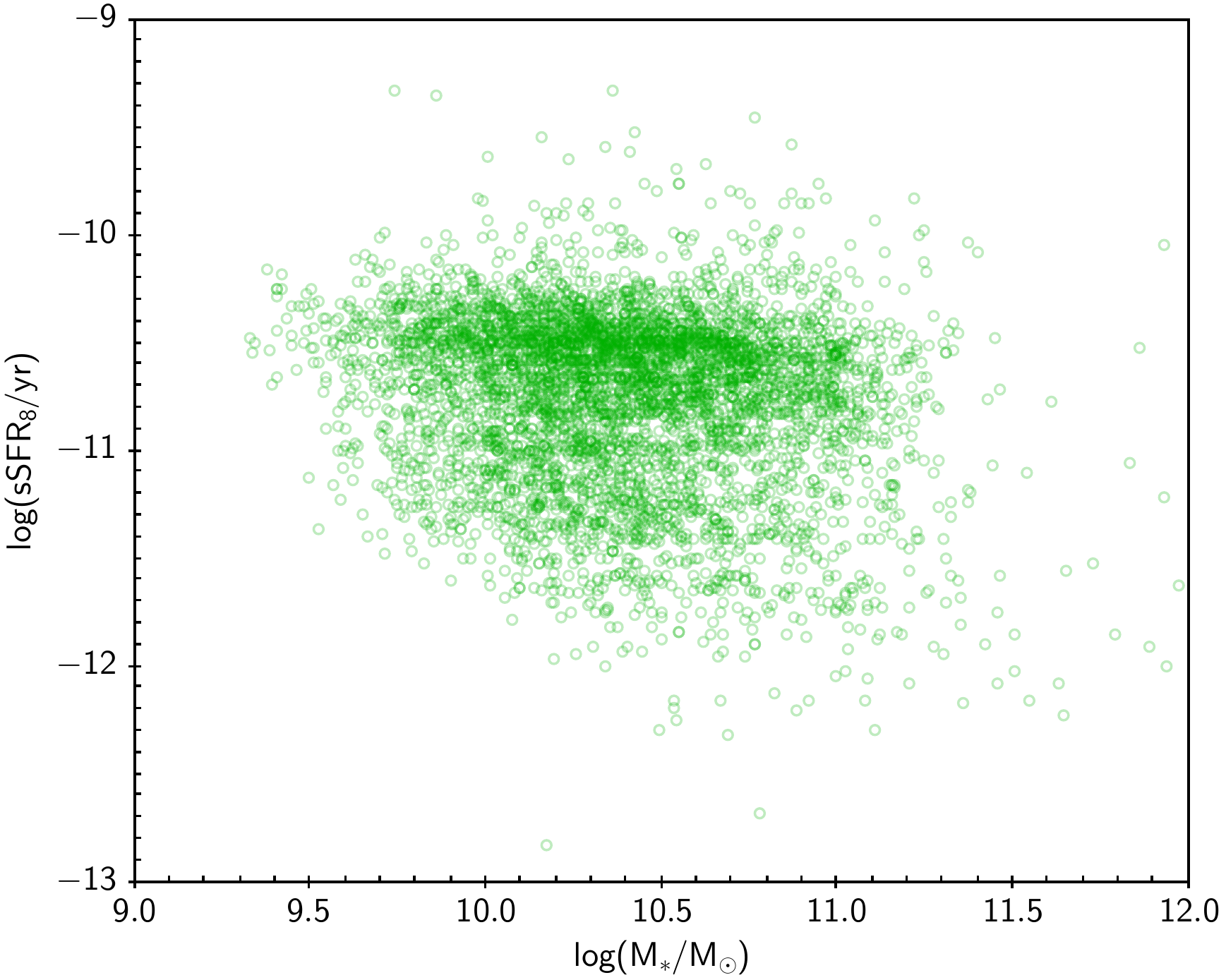}
\caption{Distribution of specific star formation rate (defined as in Fig. 3) as a function of stellar mass, for the green galaxies only. Note the peak at around sSFR$_8 \simeq 10^{-10.6}$yr$^{-1}$ at all masses.}
\label{time_ssfr_green}
\end{figure}

For orientation, we note that the majority of our green sample have a MAGPHYS-determined sSFR averaged over the last $10^8$~yr (which we label sSFR$_8$\footnote{As MAGPHYS SFRs are based on broadband colours, including the $u$-band, they naturally reflect SF on 100~Myr timescales, rather than the more `instantaneous' SFR value from, e.g. H$\alpha$ emission; see Davies et al. (2016b).}) of 
$10^{-11} <  {\rm sSFR}_8 < 10^{-10.2}$~yr$^{-1}$ (so, e.g., for a green galaxy of observed stellar mass $10^{10.6} M_{\odot}$ we have a SFR $\simeq 1 M_{\odot}$/yr), with a tail to lower values. As expected, most blue galaxies have significantly higher sSFRs (up to ${\rm sSFR}_8 \sim10^{-9.4}$yr$^{-1}$) and the red galaxies mostly have ${\rm sSFR}_8 < 10^{-10.8}$yr$^{-1}$ (see Fig. \ref{time_ssfr}). In addition, we can see that the range of green valley sSFRs is largely independent of mass (Fig. \ref{time_ssfr_green}). The overall (bimodal\footnote{The peak for the red gaxies is likely to be due, at least partly, to the estimated sSFRs being inaccurate at the low SFR end and piling up at a low but non-zero value. See the discussion in \cite{Eales2017, Eales2018} and \cite{Oemler2017} concerning the possible lack of a `valley' in the sSFR versus mass plane.}) distribution of sSFR for GAMA galaxies, and its dependence on environment, was discussed in \cite{Davies2016a}.

\section{Green Galaxy Evolution}
\subsection{Star Formation History}

The first question we should consider is whether green valley galaxies are a physically specific set of galaxies in terms of their star formation histories \citep[cf.][]{Smethurst2015} or perhaps are at a specific phase of their evolution. Despite the colour-sSFR correlation, they do not $-$ as was shown in Paper I $-$ all correspond to the same sSFR, with of order 1~dex spread at a given colour, see Fig. \ref{time_ssfr_green}. 

A straightforward indicator of the SFH is the timescale for the decline of star formation. Following common practice \citep[e.g.][et seq.]{Bell2000}, the stellar population modelling of the SEDs carried out by \cite{Taylor2011} assumes an SFR with a simple exponential fall-off, with timescale $\tau$, beginning a time $t$ before the epoch of observation (the `age' of the galaxy\footnote{It is worth noting that our `age', i.e. the time since the SFR started to decline, would be considered as the quenching timescale in some other discussions in the literature; in that sense, in our formulation galaxies have been `quenching' ever since they formed.}). The Taylor et al. modelling also assumes stellar population evolution models from \cite{Bruzual2003} with a \cite{Chabrier2003} stellar initial mass function, a variety of possible metallicities, and the Calzetti extinction law, noted earlier, with E(B-V) as another variable in the fit.

While other broad types of SFH are obviously possible, such as rising or constant SFR prior to the exponential decay \citep[e.g.][]{Lee2010,Schawinski2014,Smethurst2015,Lopez2018}, \cite{Belli2018} have shown that simple exponential SFR decays -- `tau models' -- are generally entirely adequate to characterise the SFH of most galaxies and give equivalent results to more complex models in terms of derived parameters such as the median stellar population age \citep[see also][]{Chilingarian2012}.

In addition, it is easy to see that such a SFH arises naturally in simple `closed box' models of galaxies with a Kennicutt-Schmidt \citep{Schmidt1959,Kennicutt1989} type star formation rate proportional to gas content or, indeed, similar but more general models with inflow and outflow proportional to the SFR (e.g. Tinsley \& Larson 1978; see Trussler et al. 2018 for a recent discussion in the context of quenching).

Obviously, the `tau model' does not take account of any subtleties in the SFH such as late time bursts of SF, though this is expected to be significant mainly in low mass galaxies below our sample limit.

The modelling also assumes that the whole galaxy can be represented by a single SFH, when in fact in a galaxy containing both a bulge and a disk one might expect two distinct stellar populations formed on different timescales to contribute to the overall colours \citep[e.g.][and references therein]{Lagos2018,Lopez2018}. Finally, as is well known, any parameterisation based on observed optical colours will, for strongly star forming galaxies, be weighted towards recently formed stellar populations. Luminosity weighted `ages', in particular, will be less than the actual time over which stars have formed. Nevertheless, the adopted model fits give consistent parameterisations of SFHs suitable for the present analysis.

\subsubsection{Ages and Star Formation Timescales}

\begin{figure}
\includegraphics[width=\linewidth]{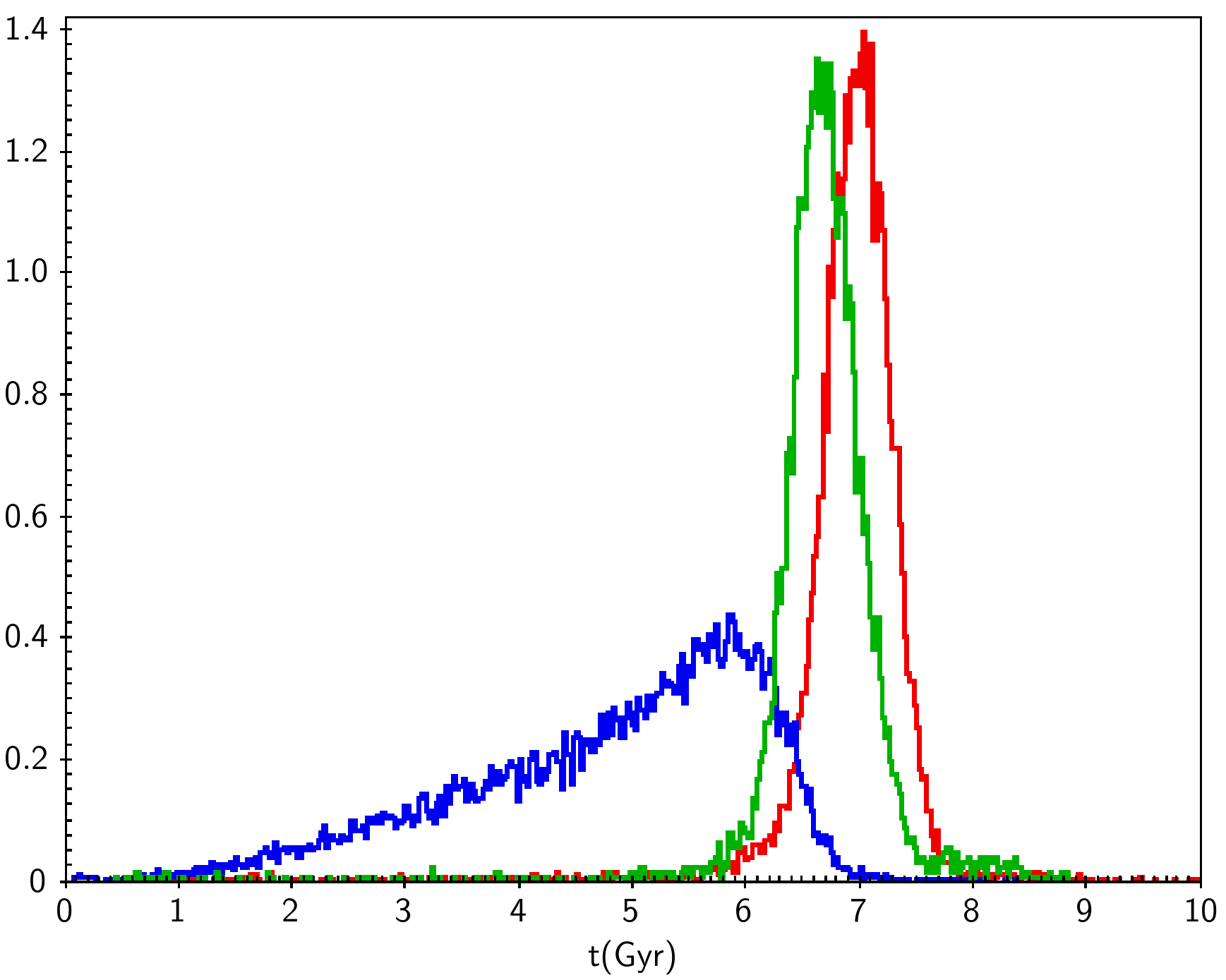}
\caption{Normalised histograms of fitted ages, $t$, for each colour subsample. Note the strong peak at around 7~Gyr for both red and green galaxies. There is very little overlap between the red and blue samples.}
\label{time_age_hist}
\end{figure}

\begin{figure}
\includegraphics[width=\linewidth]{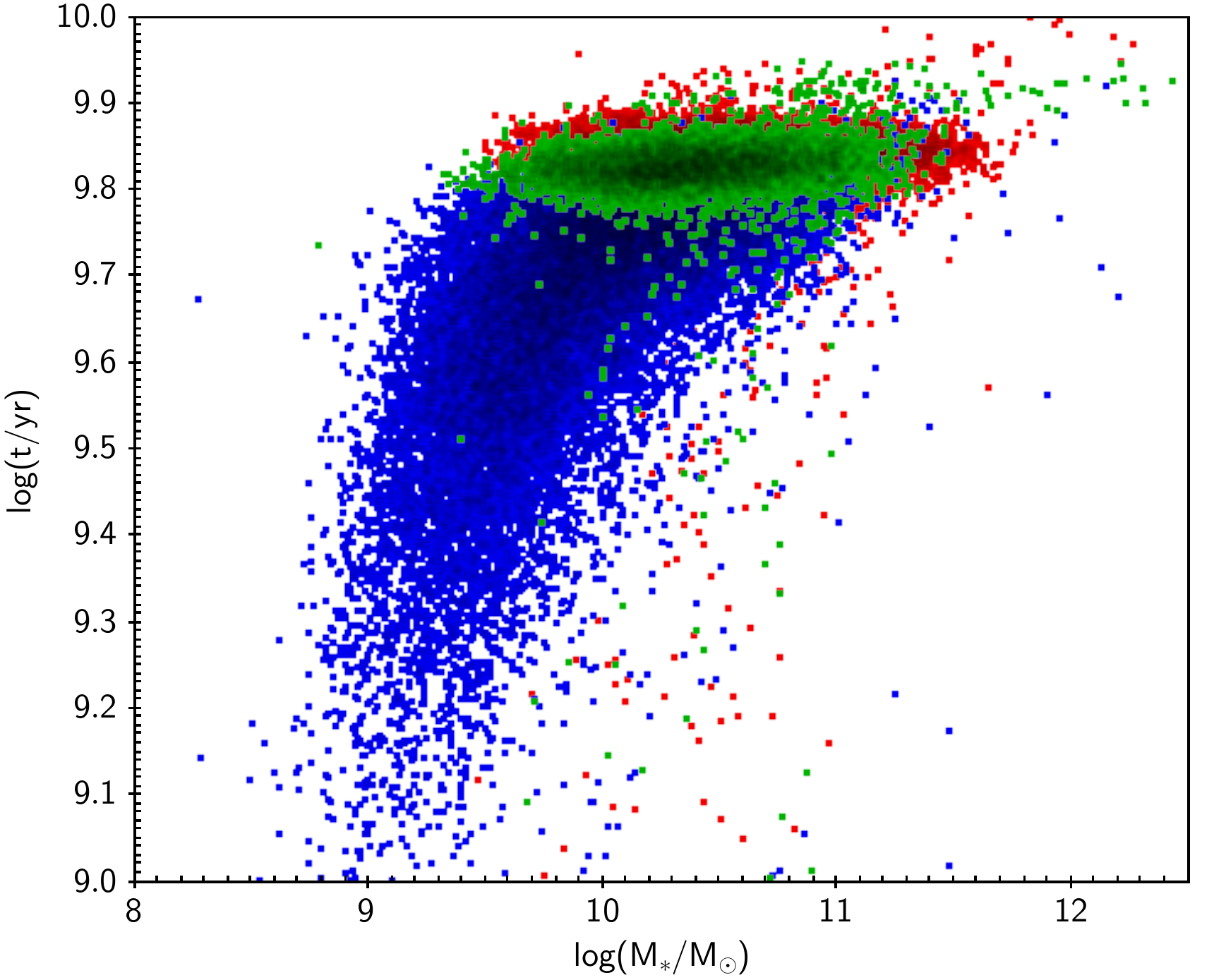}
\caption{Distribution of age, $t$, as a function of stellar mass for each colour subsample. Note the independence of age on mass for the red and green samples. Blue galaxies, on the other hand, are typically younger at lower masses (`downsizing').}
\label{time_age_mass}
\end{figure}

We note first (Fig. \ref{time_age_hist}) that there is only a small difference in age between green and red galaxies. Green galaxies did not form the bulk of their stars much more recently than reds; in both subsamples most galaxies are modelled as starting -- or, more realistically, peaking -- their star formation $\sim 7$~Gyr before the epoch of observation, i.e. $\sim 9$~Gyr ago ($z \sim 1.5$), again essentially independently of mass (Fig. \ref{time_age_mass}). Unsurprisingly, given that recent star formation tends to dominate the colours, most blue galaxies are modelled as having significantly smaller ages (even at the same mass), and the youngest galaxies are those of the lowest mass. 

The errors on individual galaxy ages, as obtained from the Bayesian analysis, are quite large, typically of order 0.16 dex in log(age), but the uncertainties are largely systematic. (The same is true for the timescales $\tau$). Changing the assumed priors can change the actual values, but largely preserves the rank order of the ages (and other parameters), so we can be confident that even with the errors, green galaxies are clearly older than blue ones, for instance. 

\begin{figure}
\includegraphics[width=\linewidth]{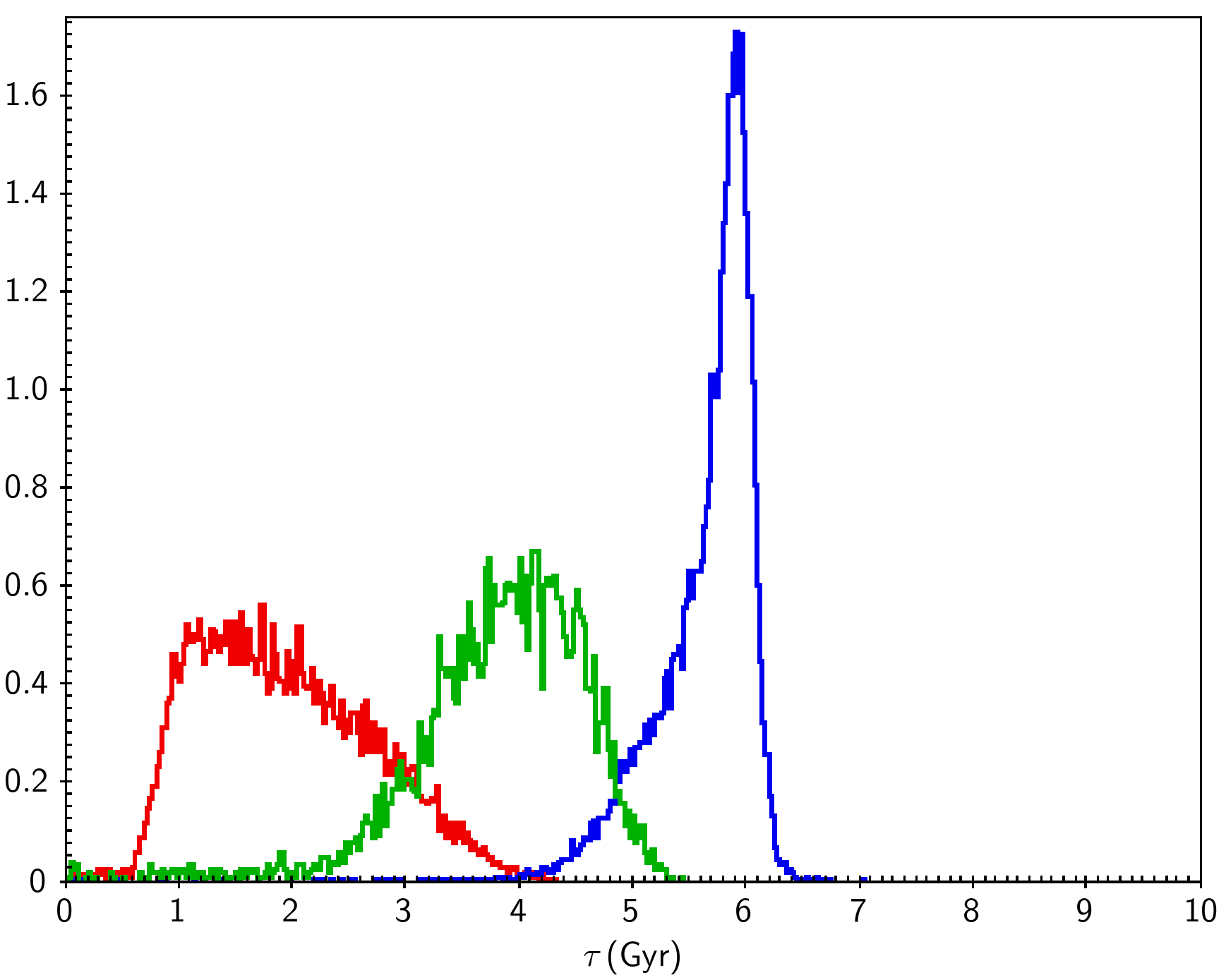}
\caption{Normalised histogram of fitted e-folding times, $\tau$, for each colour subsample. Green galaxies are clearly at intermediate values compared to the other subsamples.}
\label{time_tau_hist}
\end{figure}

\begin{figure}
\includegraphics[width=\linewidth]{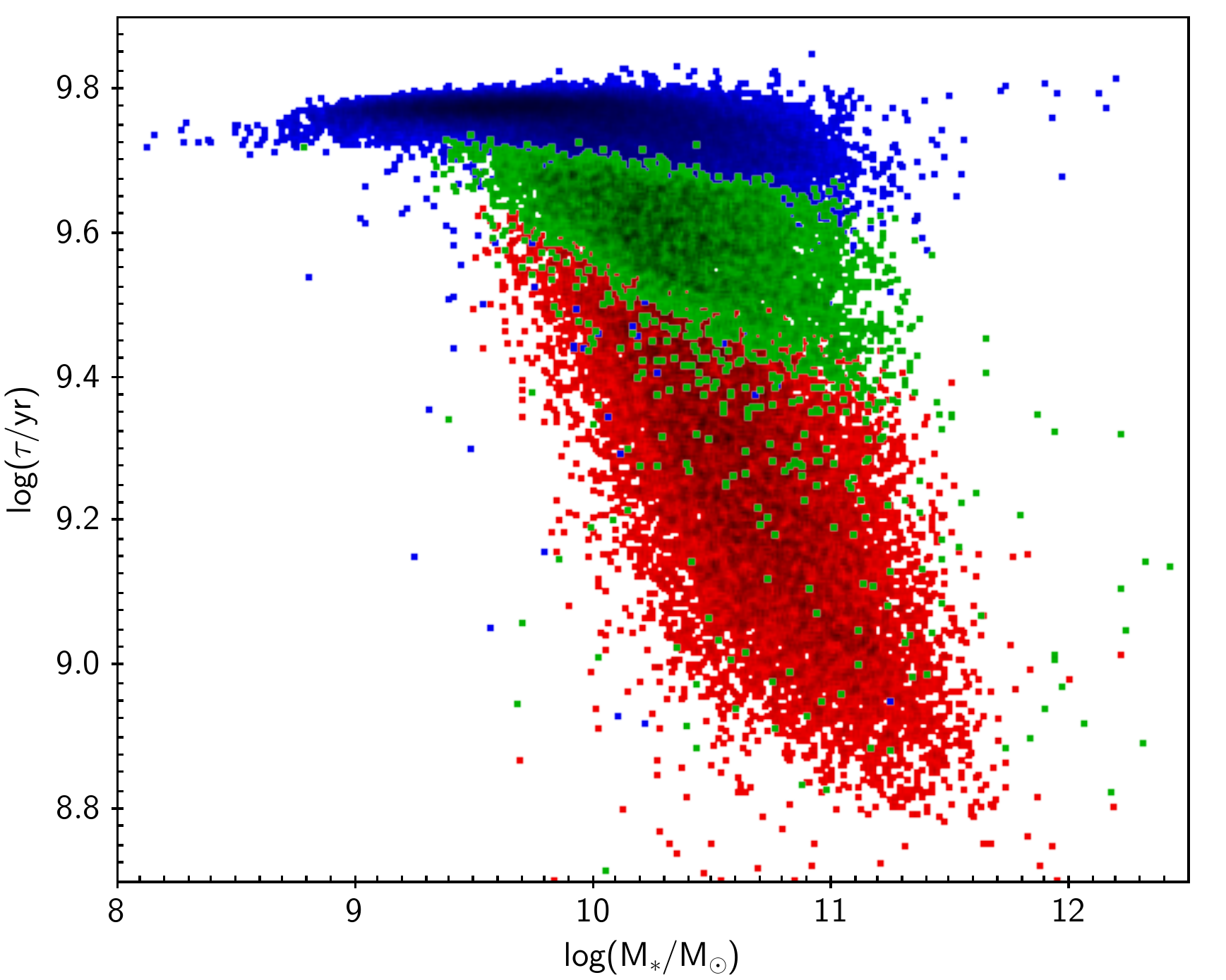}
\caption{Distribution of fitted star formation e-folding times, $\tau$, as a function of stellar mass for each colour subsample. Note the significant dependence of $\tau$ on mass for the red and green samples, while blue galaxies have similar (large) $\tau$ at all masses.}
\label{time_mass_tau}
\end{figure}

Fig. \ref{time_tau_hist} shows, next, the distribution of exponential timescales $\tau$ for the different colour subsamples. Now we see a clear separation, with the green galaxies generally at $\tau$ values between those of red and blue galaxies. However, it is evident that an intermediate timescale (typically $\tau \sim 4$~Gyr) cannot be the whole story for the green galaxies, not just because of the spread, but because the value (and spread) is clearly mass dependent (Fig. \ref{time_mass_tau}). Lower mass galaxies (of all colours) tend towards longer star forming timescales, presumably a reflection in our data and parameter space of `downsizing', i.e. ongoing star formation at recent times being preferentially seen in lower mass galaxies \citep{Cowie1996}. Even the blue galaxies have a slightly wider range of $\tau$ as mass increases. 

By looking at a sample of `green' galaxies (in GALEX plus SDSS colours) from Galaxy Zoo, \cite{Smethurst2015} found SFR e-folding times typically $\sim 2$~Gyr. These values are evidently somewhat shorter than ours from \cite{Taylor2011}. However, the two estimates cannot be directly compared, as Smethurst et al. assume a constant SFR for several Gyr prior to the exponential decline; clearly a long constant period followed by a rapid decline would be modelled as a slower overall decline in our method, so we would in general expect our $\tau$ to be longer than theirs, as is the case.

On the other hand, from a detailed study of a relatively small number of galaxies, with a range of colours and types, from the CALIFA survey, \cite{Lopez2018} find that their summed stellar populations (so effectively a galaxy of intermediate colour) can be modelled via a SF e-folding time of 3.9~Gyr, entirely consistent with our values here. They also find shorter times for early type galaxies (effectively our reds) and longer times for late type galaxies (our blues). They further note that the overall fall-off is very similar to the rate of decline of the cosmic star formation density (the cosmic star formation history) as measured by, e.g., \cite{Madau2014} and \cite{Driver2018}.

\begin{figure}
\includegraphics[width=\linewidth]{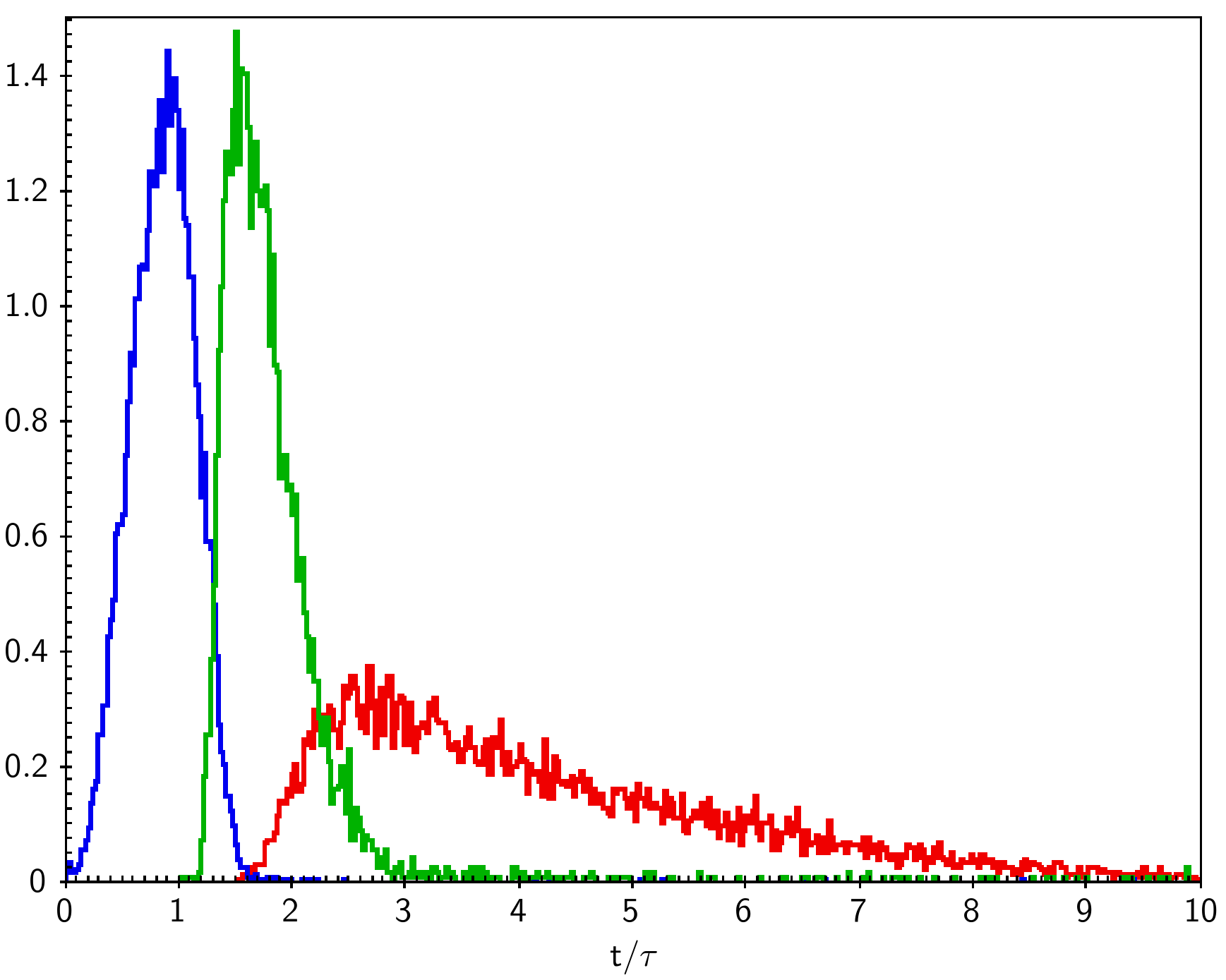}
\caption{Normalised histogram of fitted ages in units of e-folding time, $t/\tau$, for each of the colour subsamples. The green galaxies show a distinctly different distribution from the others, with values concentrated around $t/\tau \simeq 1.6$.}
\label{time_t_tau_hist}
\end{figure}

\begin{figure}
\includegraphics[width=\linewidth]{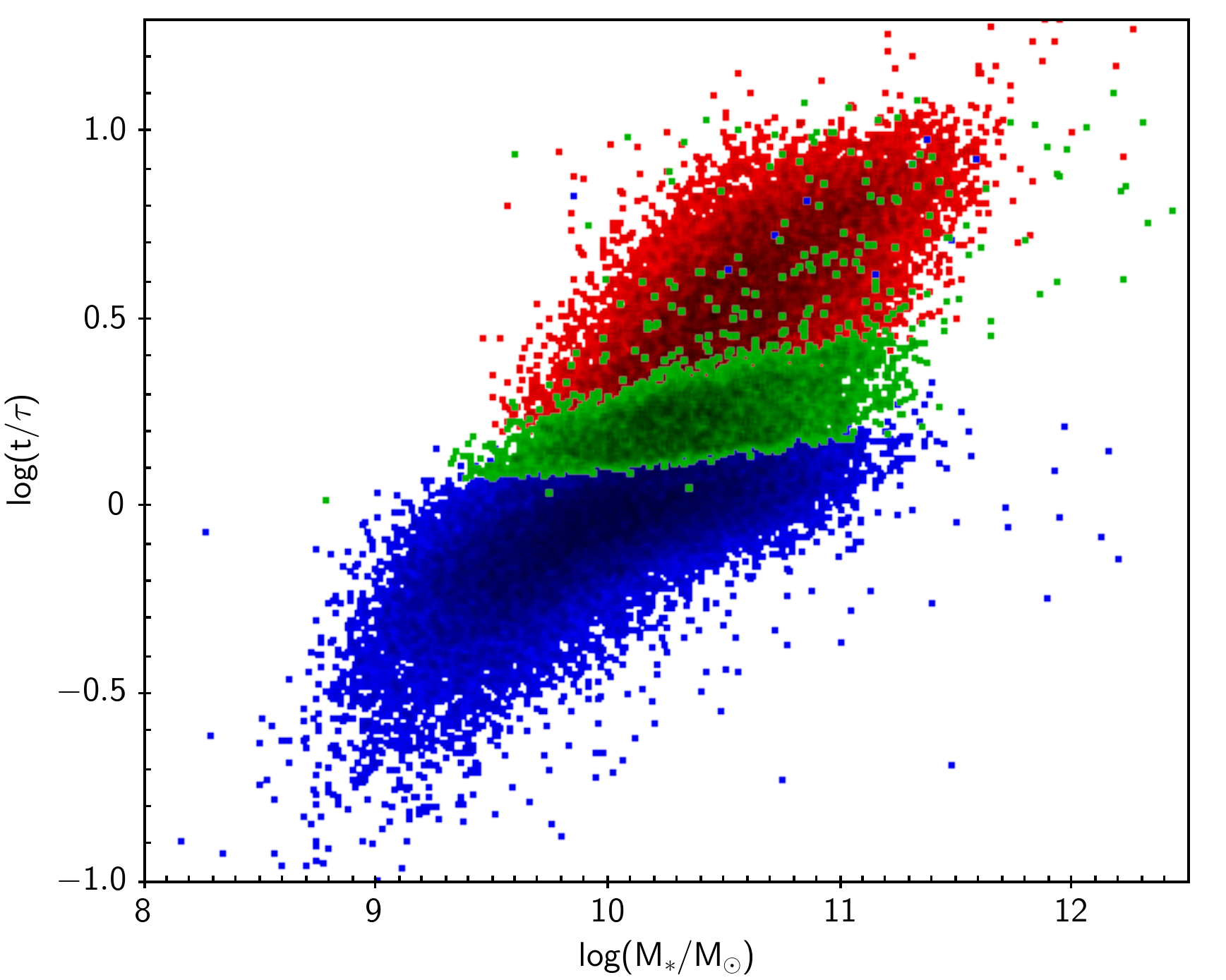}
\caption{Distribution of fitted ages in units of  e-folding times, $t/\tau$, as a function of stellar mass for each colour subsample. While green galaxies all have closely similar values, there is nevertheless a slight increase with mass, i.e. higher mass green galaxies are further through their evolution. This is also true (more strongly) for the red and blue samples.}
\label{time_t_tau_mass}
\end{figure}

Finally for this section, we consider the ratio of $t$ to $\tau$, i.e. the number of star formation e-folding times which have passed since the peak in star formation. Fig. \ref{time_t_tau_hist} shows clearly that green galaxies are intermediate between red and blue in this measure of evolution, and moreover that all green galaxies have very similar values of this parameter, with a tight concentration of values around $t/\tau = 1.6$). The distribution is significantly narrower than that in sSFR itself from Fig. 3. Nevertheless Fig. \ref{time_t_tau_mass} shows that there is some remaining small mass dependence in $t/\tau$, with a tendency towards slightly higher values (further advanced evolution) for the more massive green galaxies. This presumably arises from the somewhat redder colours selected for the green valley at higher mass. It could possibly also reflect an environment dependence, as more massive galaxies are typically in larger groups. We defer any further discussion of the environment of green galaxies to a subsequent paper.

Of course, a correspondence between a colour subset and the evolutionary state is not unexpected, given the \cite{Bruzual2003} models used throughout our analysis. However, there was no a priori necessity for the green valley galaxies to be specifically those galaxies fitted by an old age $t$ but an intermediate timescale $\tau$ and with a narrow distribution of $t/\tau$.  

\subsection{Colour Evolution}
The above provides, at least qualitatively, the answer to our first question, i.e. it does appear that green galaxies are at a specific characteristic point in their evolution, where the star formation has decreased by a factor $\simeq 5$ relative to its peak value.

\begin{figure}
\includegraphics[width=\linewidth]{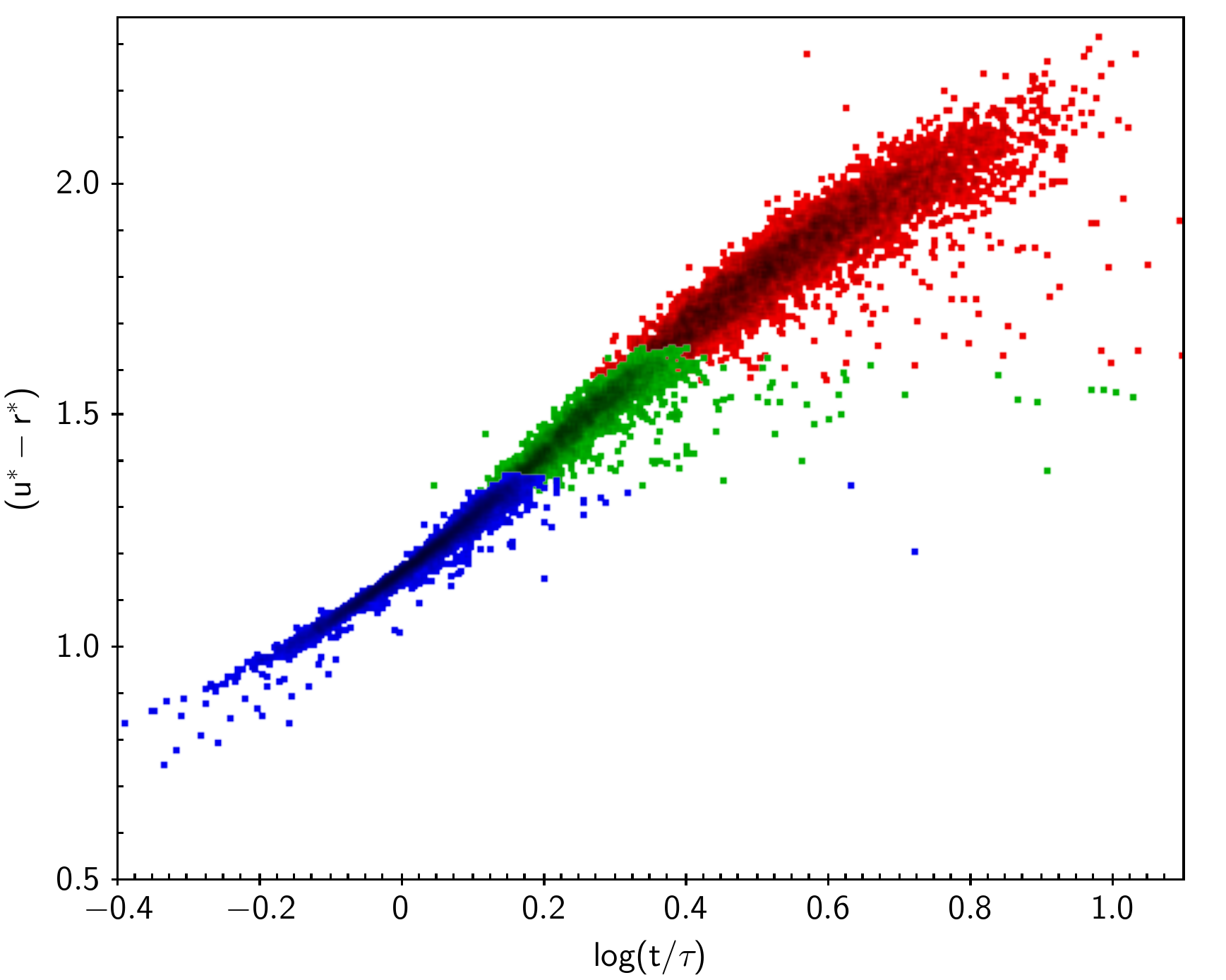}
\includegraphics[width=\linewidth]{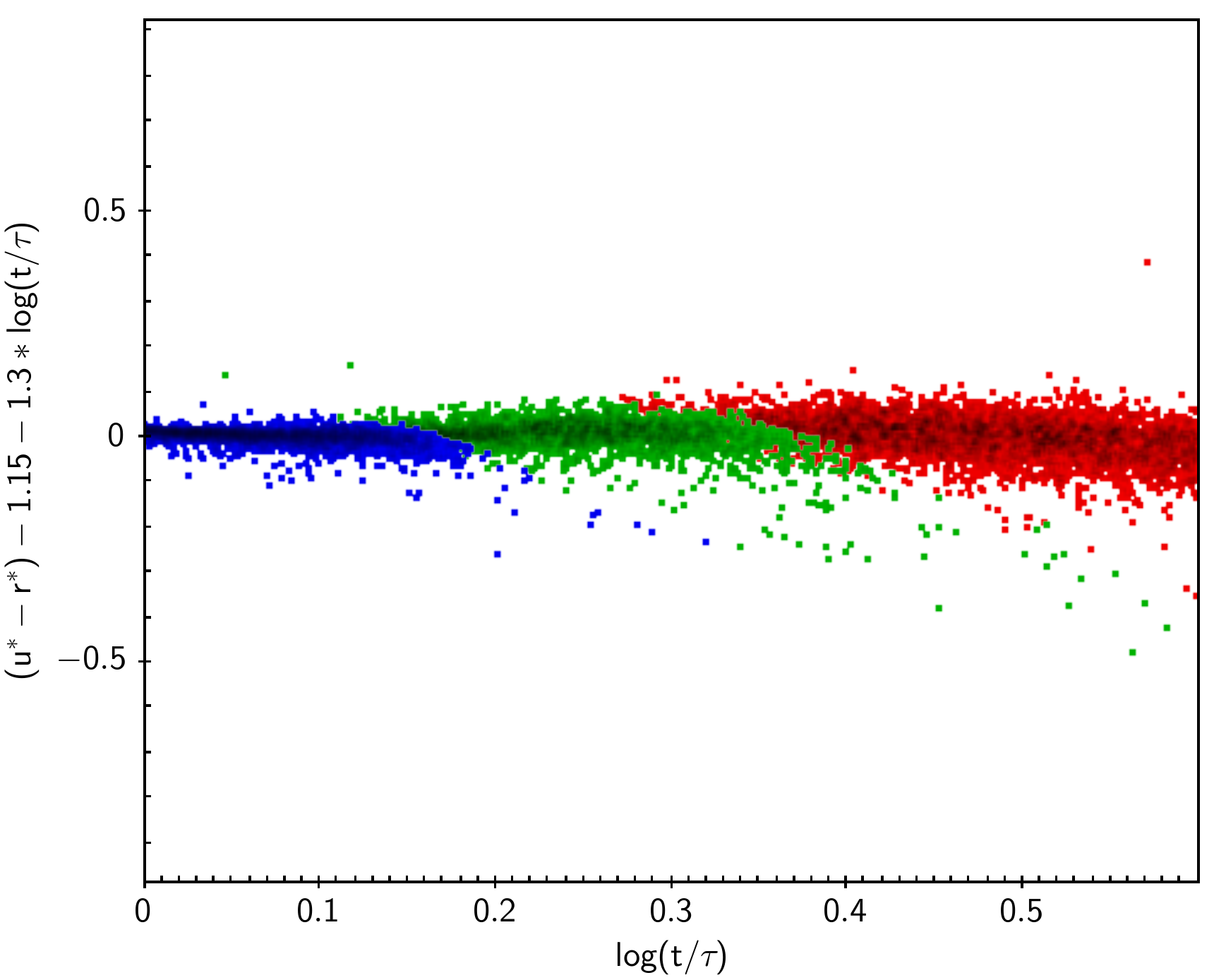}
\caption{Variation of ($u^* - r^*$) colour against fitted age in units of e-folding times, $t/\tau$ (top), and relative to the overall correlation (bottom), for the mass limited subsample.}
\label{time_pseudocolour_mass10_figure}
\end{figure}

Indeed, given the similarity between Fig. \ref{time_mass_colour} and Fig. \ref{time_t_tau_mass}, and the well constrained distribution of $t/\tau$ for green galaxies, we should not be surprised that there is a strong correlation between $t/\tau$ and ($u^* - r^*$), as demonstrated, for example, for the mass limited subsample in the top panel of Fig. \ref{time_pseudocolour_mass10_figure}.  The best fit straight line correlation across the range $t/\tau$ between 1 and 4 is given by 
\[(u^* - r^*) = 1.15 + 1.3 \: {\rm log}(t/\tau).\]
The bottom panel of  Fig. \ref{time_pseudocolour_mass10_figure} demonstrates the small scatter relative to this best fit line, which also holds true for the whole main sample; irrespective of mass the scatter is less than $\pm 0.1$ magnitudes

Notice that this implies that as galaxies of any given $\tau$ age, they move {\it along} the same overall relation. This is evident from the bottom panel of Fig. \ref{time_pseudocolour_mass10_figure}, since the different colour subsamples have different typical values of $\tau$ (and is easily checked from subsamples of fixed $\tau$). 

The close correlation then implies that we can empirically estimate how the colour should evolve over time, viz. a change by $\simeq 0.4$ magnitudes as $t/\tau$ changes by a factor 2. We should note here that this correlation {\em only} applies in the plotted range where colour is an essentially linear function of evolutionary state. Outside this range the colours vary less with evolutionary state (bending to a slope of $\simeq 0.8$ rather than the 1.3 in the above equation; see the top panel in Fig. \ref{time_pseudocolour_mass10_figure}) as they asymptote to those of totally star forming or totally dead stellar populations \citep[cf.][]{Oemler2017, Eales2017}. 

\begin{figure}
\includegraphics[width=\linewidth]{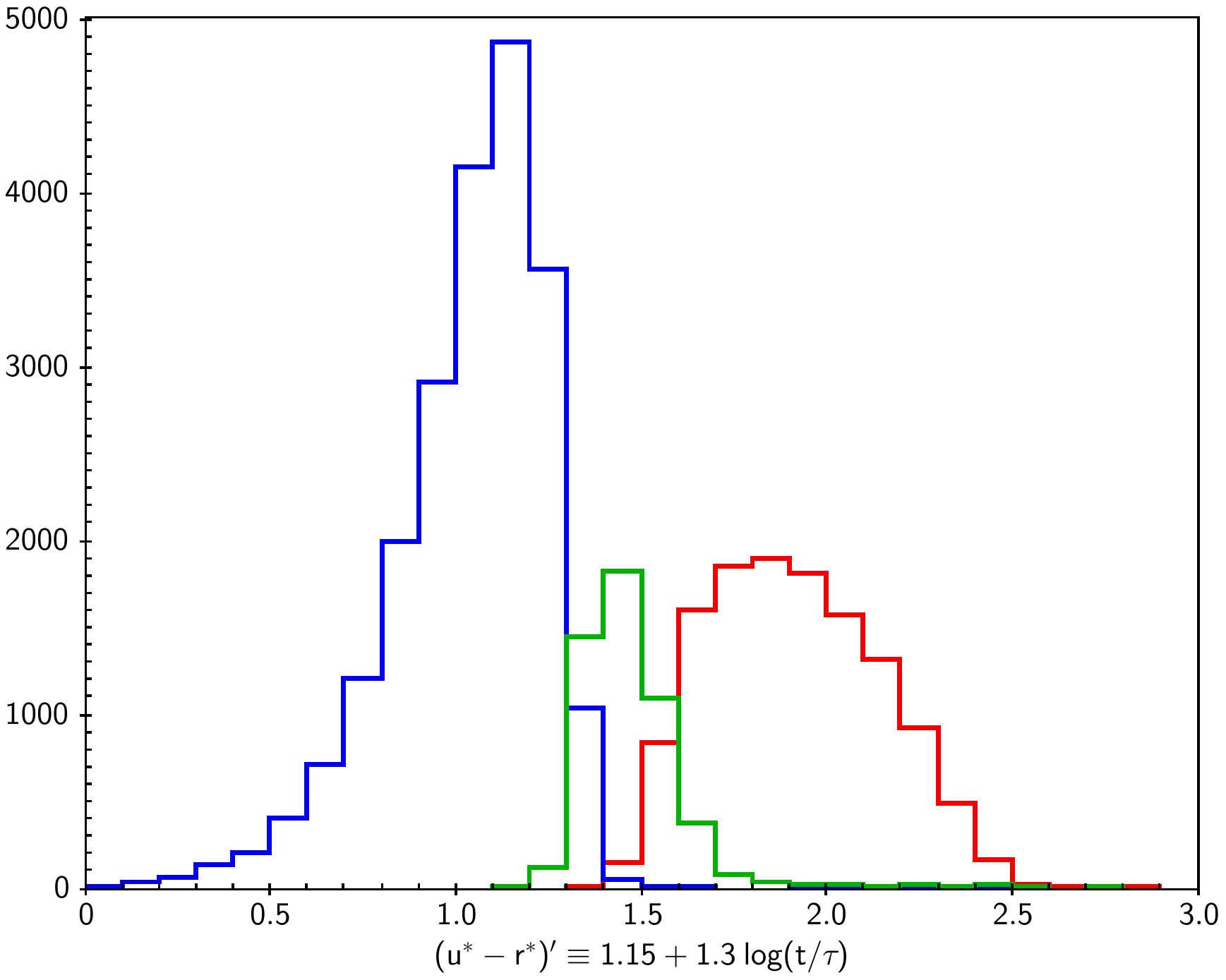}
\caption{Histograms of the pseudo-colour $(u^* - r^*)'$ (see text) for each colour subsample of our main sample. Note the strong resemblance to Fig. \ref{time_colour_hist}, showing the actual $(u^* - r^*)$ colours.}
\label{time_synthcolour_hist}
\end{figure}

Fig. \ref{time_synthcolour_hist} shows the distribution of the `pseudo-colour' $(u^*~-~r^*)'$, obtained from $t/\tau$ via the above relation, i.e.
\[(u^* - r^*)' \equiv 1.15 + 1.3 \: {\rm log}(t/\tau),\]
for each of our colour subsamples. We can note the strong resemblance to the original (actual) colour distribution in Fig. \ref{time_colour_hist}. We therefore use the empirically predicted evolution of this pseudo-colour as a proxy for the evolution of the real colour. We return to this below.

\subsection{Specific Star Formation Rates}

Before we do this, though, to get a rough first estimate of the timescales involved, we consider the expected evolution of the sSFR for `typical' galaxies. From Fig. \ref{time_ssfr}, we can say that the typical red galaxy has log(sSFR) $ \simeq -11.5$. A typical green galaxy, on the other hand, has log(sSFR) $\simeq -10.6$. To reach the middle of the red sequence a `typical green' therefore has to reduce its SFR (in the absence of any significant change in stellar mass) by a factor $\sim 8$. Further, from Fig. \ref{time_tau_hist}, green galaxies have $\tau \sim 4$~Gyr. Matching this to the assumed SFR decline $e^{-\Delta t/\tau}$, then $\Delta t/\tau \simeq $ ln$\: 8$, i.e. $\Delta t \sim 8$~Gyr. Thus despite various other similarities, noted here (age), in Paper I (S\'{e}rsic index, bulge-to-disk mass ratio) and elsewhere \citep[e.g.][for surface density]{Fang2013}, it is evident that in global star formation terms green valley galaxies (with fitted ages $\sim 7$~Gyr) are only `half way' to becoming a thoroughly `red and dead' early type (unless their SFR decline speeds up, of course). 

However, this is not the same as saying that galaxies stay green for this length of time. Leaving the green valley merely requires them to cross the border into red territory.

If we take, as a specific example, galaxies `in the middle' of the green valley to be within 0.02 magnitudes of the central colour (for their mass), then we can check that, similar to what was assumed above, they have a mean log(sSFR) $\simeq -10.7$ (and a mean $\tau$ of 4.0~Gyr, as for the whole green sample). If we similarly take the subset of galaxies redder than the edge of our green valley by less than 0.02 magnitudes\footnote{The size of our large GAMA data set allows such detailed subdivision, each of the sub-subsamples we use here has $\sim 500$ galaxies.}, we find that they have an average log(sSFR) $\simeq -11.1$, i.e. a factor 2.5 lower.  Thus a typical green will take $\sim 3-4$~Gyr to (just) become a red galaxy or, rather, to reach the same sSFR as a current red galaxy (we consider actual colour evolution in the next section). We can think of this as an estimate of the `half-life' of a current green galaxy; after this time the redder/lower sSFR half of them (at a given mass) will have evolved sufficiently to move out of the typical range of sSFRs observed for current green valley galaxies. Note that for simplicity, throughout the paper we will refer to our $0.1<z<0.2$ samples as the ``current'' green valley, etc., to distinguish this from higher redshift studies, though strictly the galaxies are observed as they were 1-2 Gyr ago. 

We can extend this to estimate how long a galaxy at the blue side of the green valley will take to cross to the red side. Galaxies within 0.02 magnitudes of the blue edge of the green valley have a mean log(sSFR) $\simeq -10.5$ and mean $\tau \simeq 4.7$~Gyr. For one to reach the middle of the green valley, the sSFR must change by a factor 1.6, which will take $\Delta t \sim 2$~Gyr. Hence, to complete the crossing of the green valley from blue to red should take a total of $\sim 5-6$~Gyr. Notice that the middle of the valley is rather nearer to the blue edge than the red edge in star formation terms (factor 1.6 rather than 2.5) and this more than makes up for the slightly longer e-folding times of the bluer galaxies, making the first half of the transition faster overall. We should not necessarily believe the details of these estimates, of course, as we should really allow for the spreads in sSFR (or colour) and $\tau$. In particular, galaxies (of any colour) with smaller $\tau$ will change their sSFR and colour more quickly. In addition, as seen earlier, $\tau$ and sSFR both depend on mass and there is a sizeable scatter in sSFR at a given colour. We therefore make more detailed calculations in terms of the evolution in colour in the following section.

Moving outwards from the green valley, it is evident that most blue galaxies near the edge of the green valley have a $\tau$ significantly longer, around 5~Gyr, than typical green galaxies themselves (4~Gyr), and those further away generally have even longer $\tau \sim 6$~Gyr. This may indicate that either (i) only some fraction of blue galaxies (in the tail of the $\tau$ distribution) can turn green over a relatively short timescale  or (ii) the rate of decline of star formation speeds up ($\tau$ decreases) for some (or all) blue galaxies once their SFR reaches some threshold, perhaps due to the (un)availability of sufficiently dense fuel. 

It is worth noting that for the simple case of a closed-box model with SFR proportional to gas mass -- which results in an exponential fall-off, as assumed in our analysis -- $\tau$ is equal to the gas depletion timescale derived simply as the remaining gas mass divided by the current SFR \citep{Roberts1963}. 

Notice that point (ii) above reflects that there may be a certain inconsistency in our simple approach, in that we assume that $\tau$ remains fixed for each half of the transition (blue to green and green to red), while we know that blues have longer $\tau$ than greens which in turn have longer $\tau$ than reds, perhaps suggesting a steady shortening of the timescale while traversing the valley (though see Section 3.5 for arguments against this). Either way, the above arguments should give us a plausible first estimate without requiring us to model a more complex form for the SFH. 

With respect to numbers, if we do split the evolution into two phases, then from our calculation we can say that a galaxy in the redder (lower sSFR) half of the green valley galaxy should have escaped from the valley (in SF terms) in $3-4$~Gyr time. Over the same timescale, some current blue (higher sSFR) galaxies should enter. Looking at this purely from the sSFR (we return to the colours in the next section), the blue galaxies which can enter the green valley in the next 3~Gyr are evidently those which are sufficiently close to the green valley boundary that with their combination of sSFR and $\tau$, schematically, sSFR$_{blue} e^{-3/\tau_{blue}} \rightarrow $ sSFR$_{green}$. If the number in this `pre-transition' region is a suitable fraction of all blue galaxies then we can repopulate the green valley at an appropriate rate to keep a sensible fraction of galaxies in the green valley at all times \citep[see e.g. Paper 1 and, at higher redshifts,][]{Cooper2006,Forrest2018}. 

\subsection{Colour Evolution Revisited}

To develop this investigation further, we next use the whole distributions of $\tau$ and sSFR. On the {\it assumption} that $\tau$ for a given galaxy is fixed over the relevant timescale (and that there is negligible change in the total stellar mass), to see how the green valley will be populated at different times we can, as above,  evolve each galaxy's sSFR forward \citep[or indeed backwards; see][]{Belli2018,Lopez2018} for a time $\Delta t$ according to sSFR $\rightarrow$ sSFR $e^{- \Delta t/\tau}$, which translates to
\[{\rm log(sSFR)}(\Delta t) = {\rm log(sSFR}_{obs}) -0.4343 \: \Delta t /\tau \; . \]

Working in sSFR itself is problematical, though, as green galaxies are spread over a significant range ($\sim 1$~dex) of sSFR and at almost all sSFR values, even at the centre of the valley, there are actually more red and blue galaxies at a given sSFR than green (see Fig. \ref{time_ssfr}). We show some plots of evolved sSFRs themselves in the Appendix, but in this section and the rest of the main body of the paper we concentrate on the colours, or the pseudo-colours $(u^* - r^*)'$ from Section 3.2. 

\begin{figure}
\includegraphics[width=\linewidth]{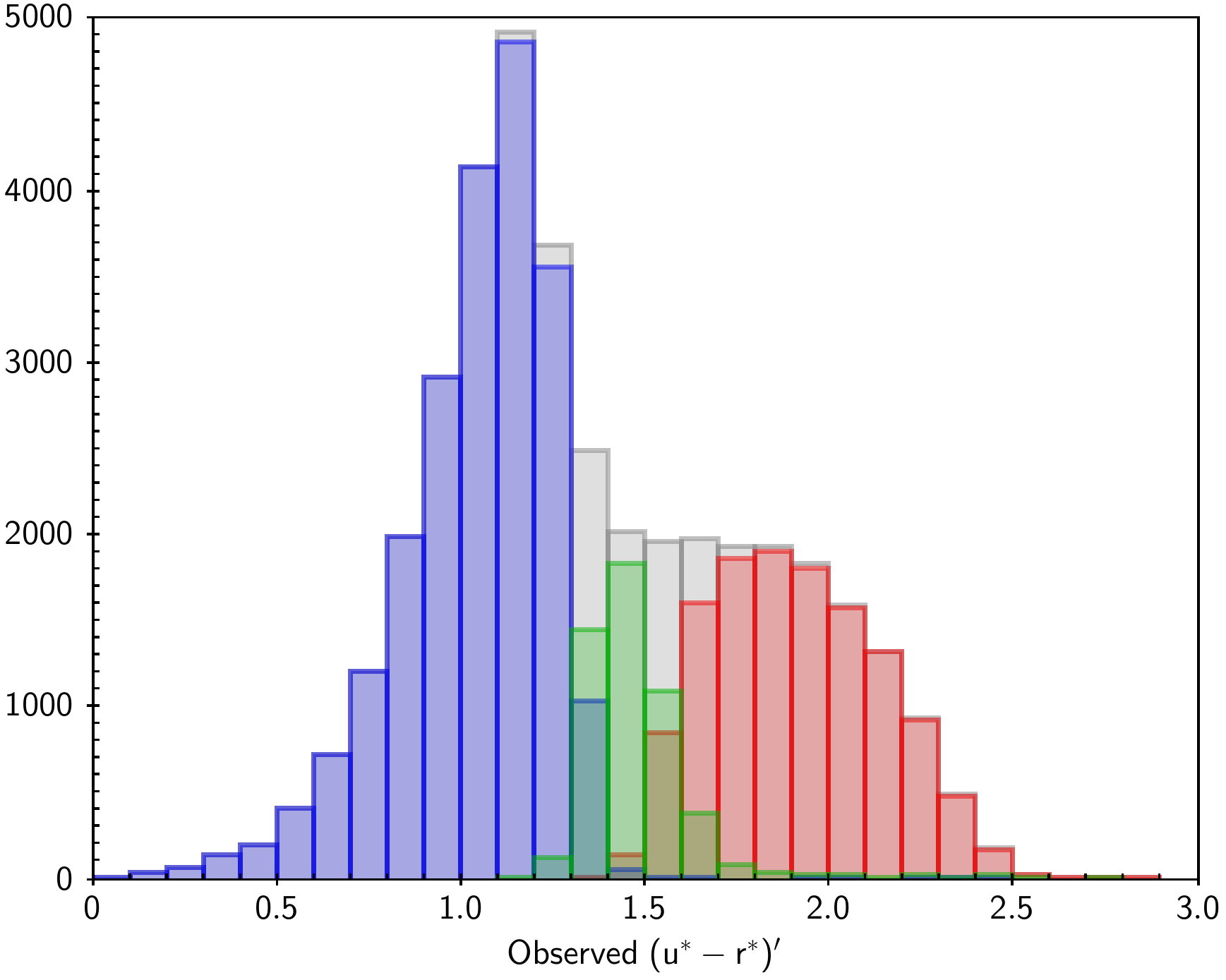}
\includegraphics[width=\linewidth]{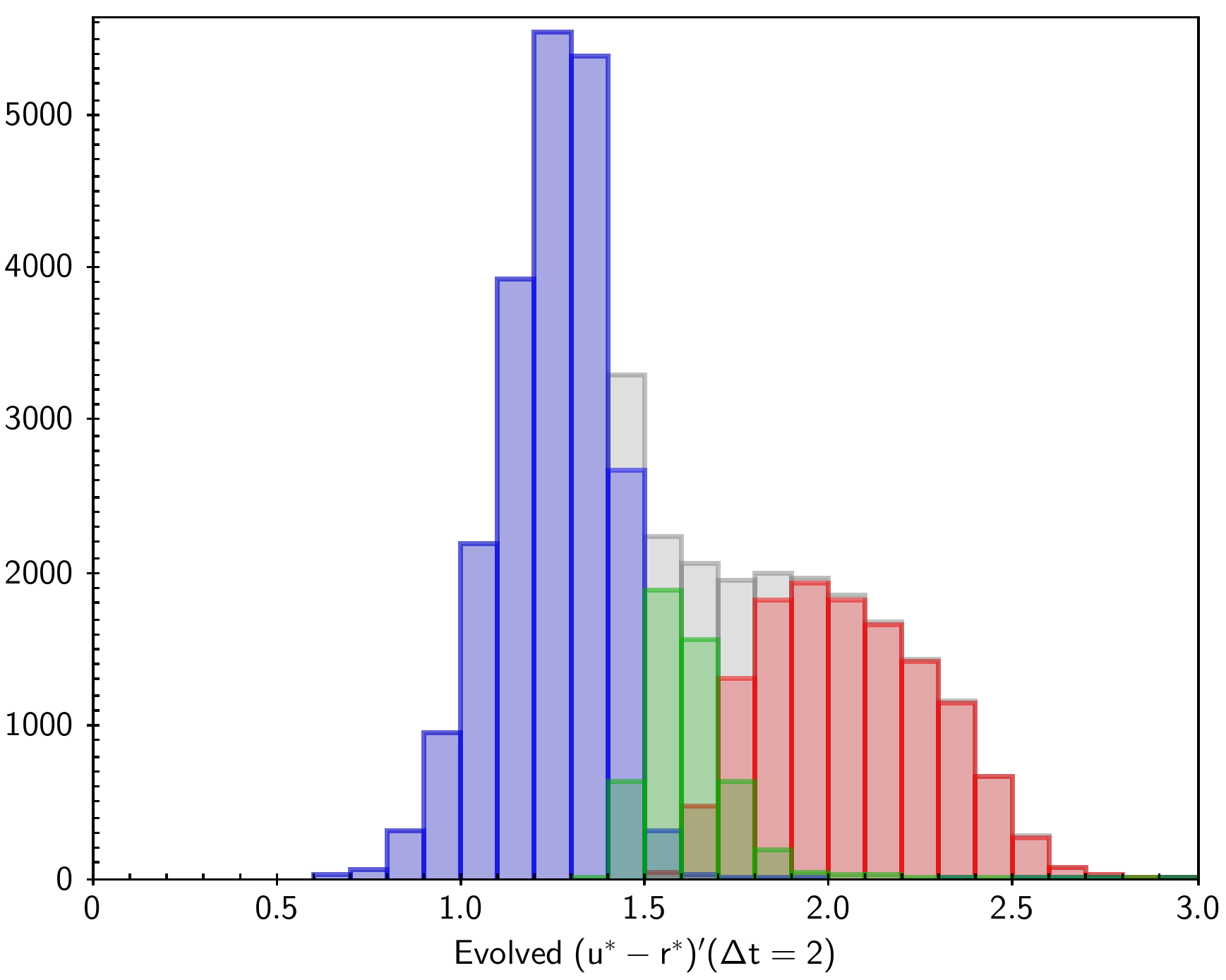}
\includegraphics[width=\linewidth]{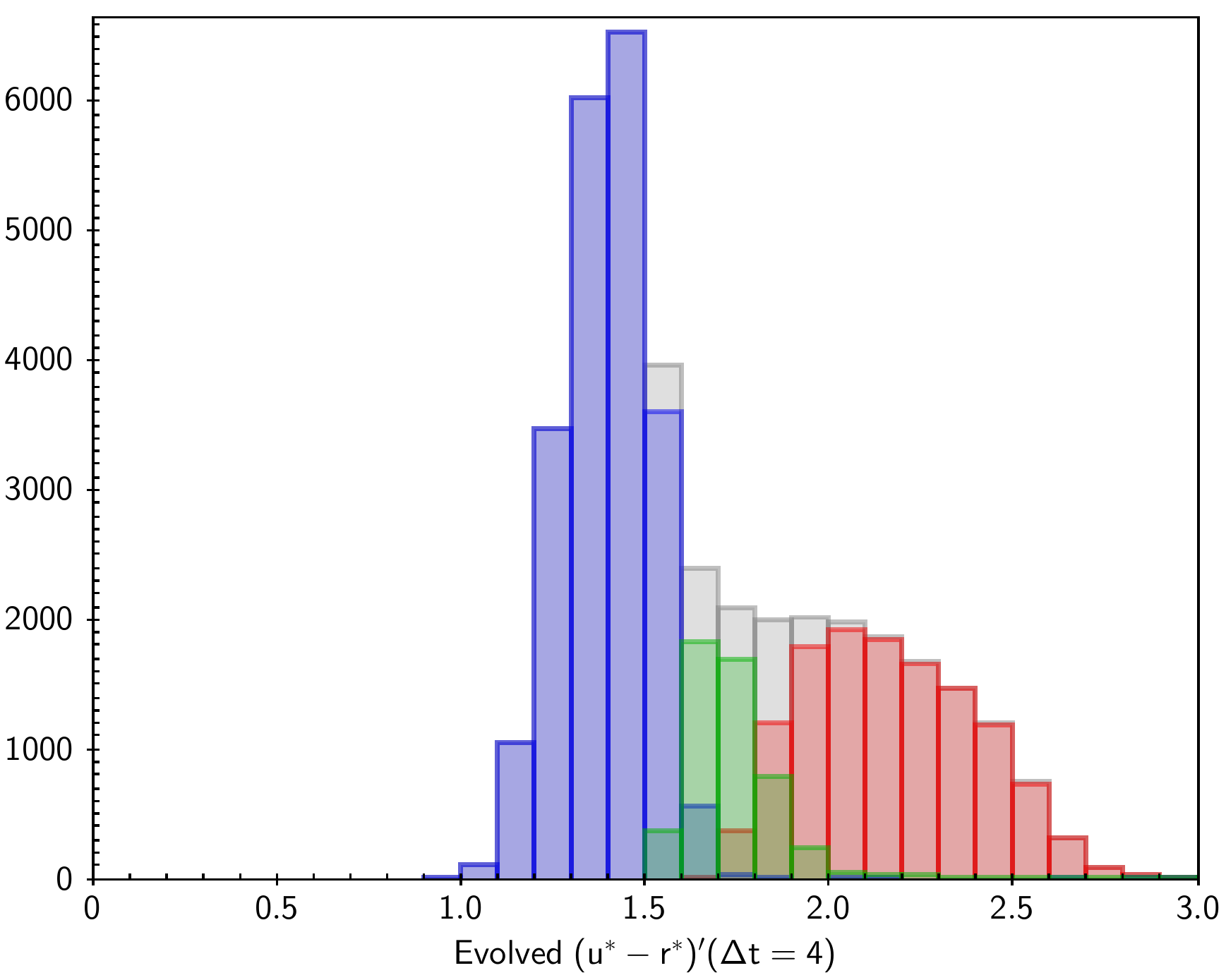}
\caption{Histograms of pseudo-colour $(u^* - r^*)'$ for the colour subsets: (top) as observed (as Fig. 12), (middle) as evolved forwards by 2~Gyr and (bottom) as evolved forwards by 4~Gyr, assuming a constant $\tau$ for each galaxy (see text). The total distributions are given by the grey histograms.}
\label{time_synthcolour_evolution}
\end{figure}

We can easily evolve the distribution of pseudo-colours by adding any chosen $\Delta t$ to the fitted ages of our galaxies and recalculating $(u^* - r^*)'$. As examples, the middle and bottom panels of Fig. \ref{time_synthcolour_evolution} show the evolved pseudo-colours for $\Delta t = 2$~Gyr and 4~Gyr, compared to the observed distribution (cf. Fig. \ref{time_synthcolour_hist}) in the top panel.

 After 2~Gyr we can see that a large number of the green galaxies have already moved redwards of $(u^* - r^*)' = 1.7$. By $\Delta t = 4$~Gyr, obviously, even more current green galaxies have  $(u^* - r^*)' > 1.7$. However, this is not exactly the definition of the edge of the green valley, as that is mass dependent.

\begin{figure}
\includegraphics[width=\linewidth]{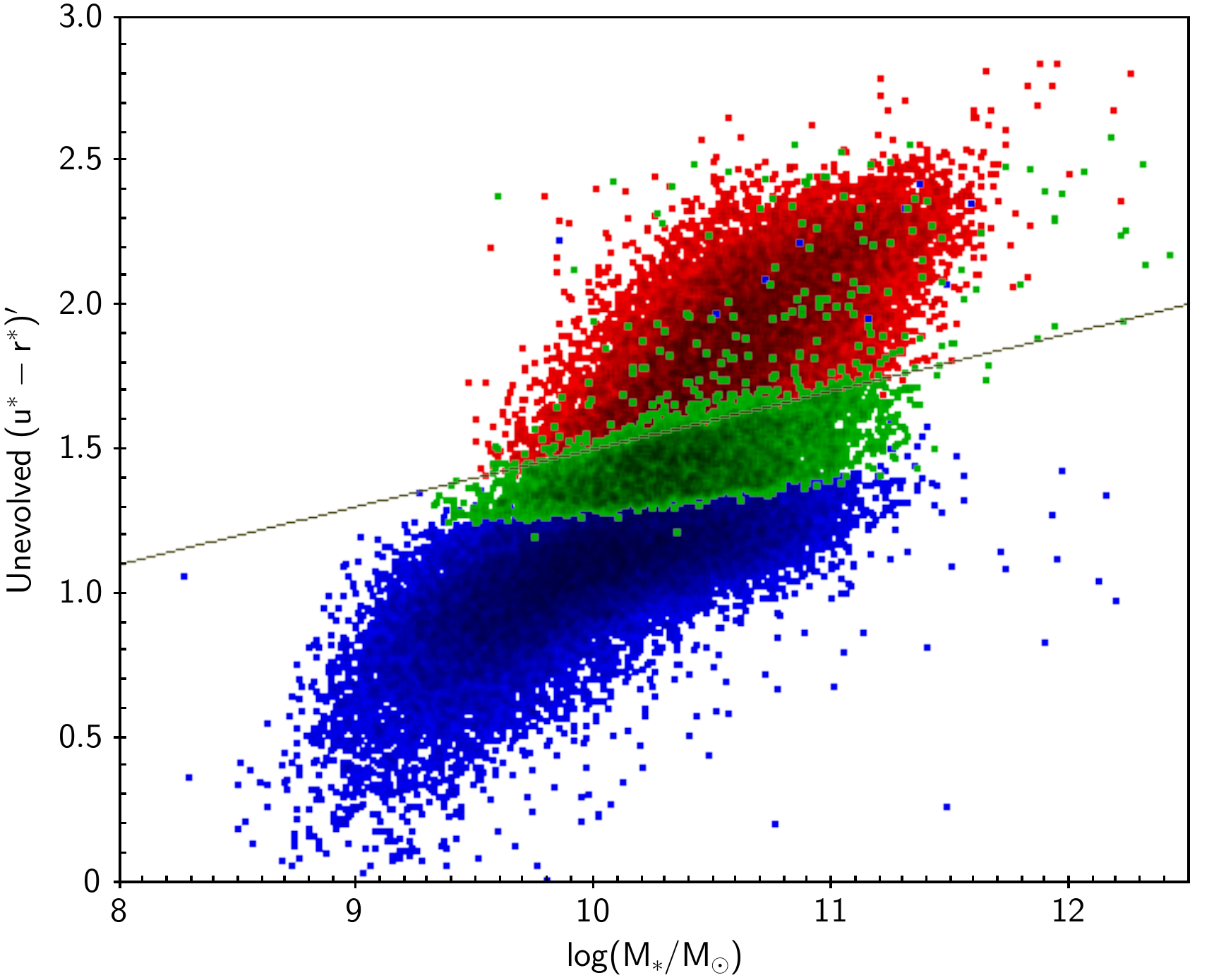}
\includegraphics[width=\linewidth]{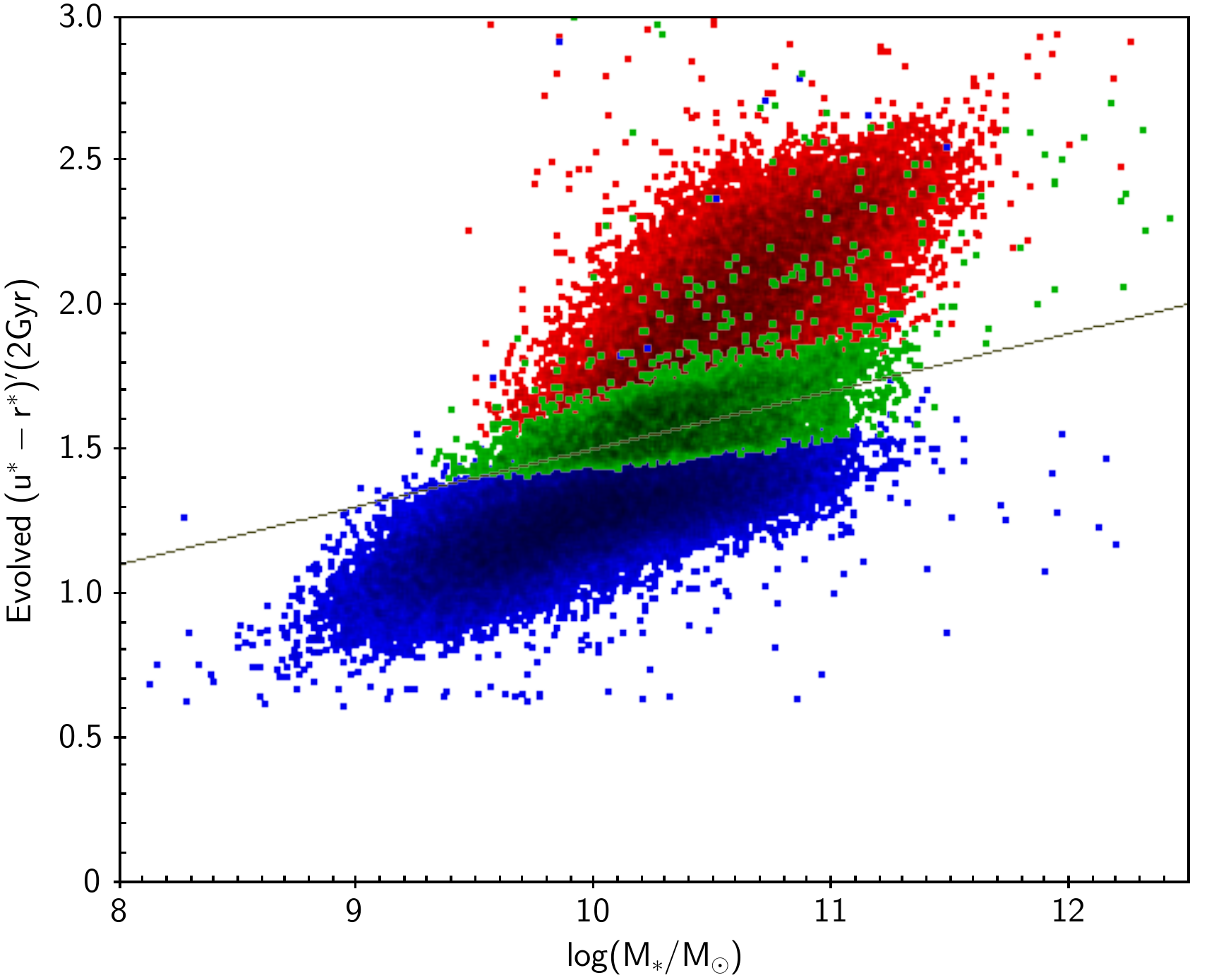}
\includegraphics[width=\linewidth]{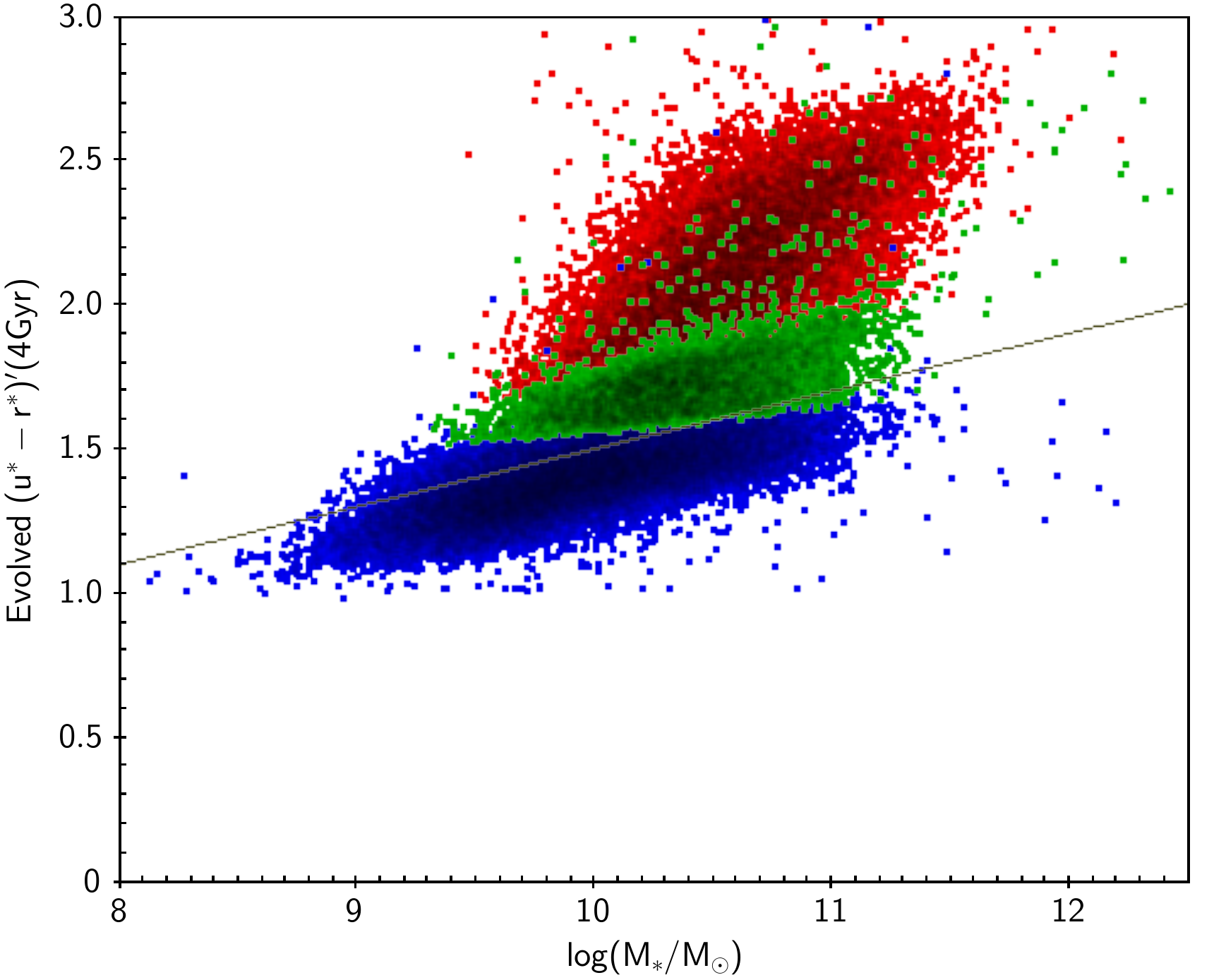}
\caption{Distribution of pseudo-colour $(u^* - r^*)'$ as a function of (observed) mass, (top) as observed, (middle) as evolved forwards by 2~Gyr and (bottom) as evolved forwards by 4~Gyr (see text). The diagonal black lines in the middle and bottom panels represent the red edge of the (true) colour green valley from Fig. 1, demonstrating that virtually all current green galaxies have evolved out of the current green valley after 4~Gyr.}
\label{time_synthcolour_mass_evolution}
\end{figure}

Therefore to get a more accurate picture, in the middle panel of Fig. \ref{time_synthcolour_mass_evolution} we now show the whole evolved pseudo-colour versus mass plane after 2~Gyr. (The top panel shows the observed distribution). Note that we use the mass at the epoch of observation in all the plots, and do not attempt to account for any, generally small, changes of stellar mass with time. Of the $\sim 5000$ green galaxies, $\sim 2800$ have crossed the (current) green valley upper border, to become red, if not yet dead. On the other hand, of course, a number of formerly blue galaxies have `moved in'. Indeed, because of the large initial population of blues, this more than compensates and we have a total of $\sim 9500$ galaxies in the area of the current green valley.

After 4~Gyr ( Fig. \ref{time_synthcolour_mass_evolution} bottom panel) all except a handful of the high mass galaxies among the current green galaxies sample have moved out of the green valley as currently defined (its top edge is indicated by the diagonal line). Indeed this is actually largely true by 3~Gyr. These (formerly) green galaxies have been replaced (in the current valley) by a large number ($\sim 15000$) of the current blue galaxies. 

\begin{figure}
\includegraphics[width=\linewidth]{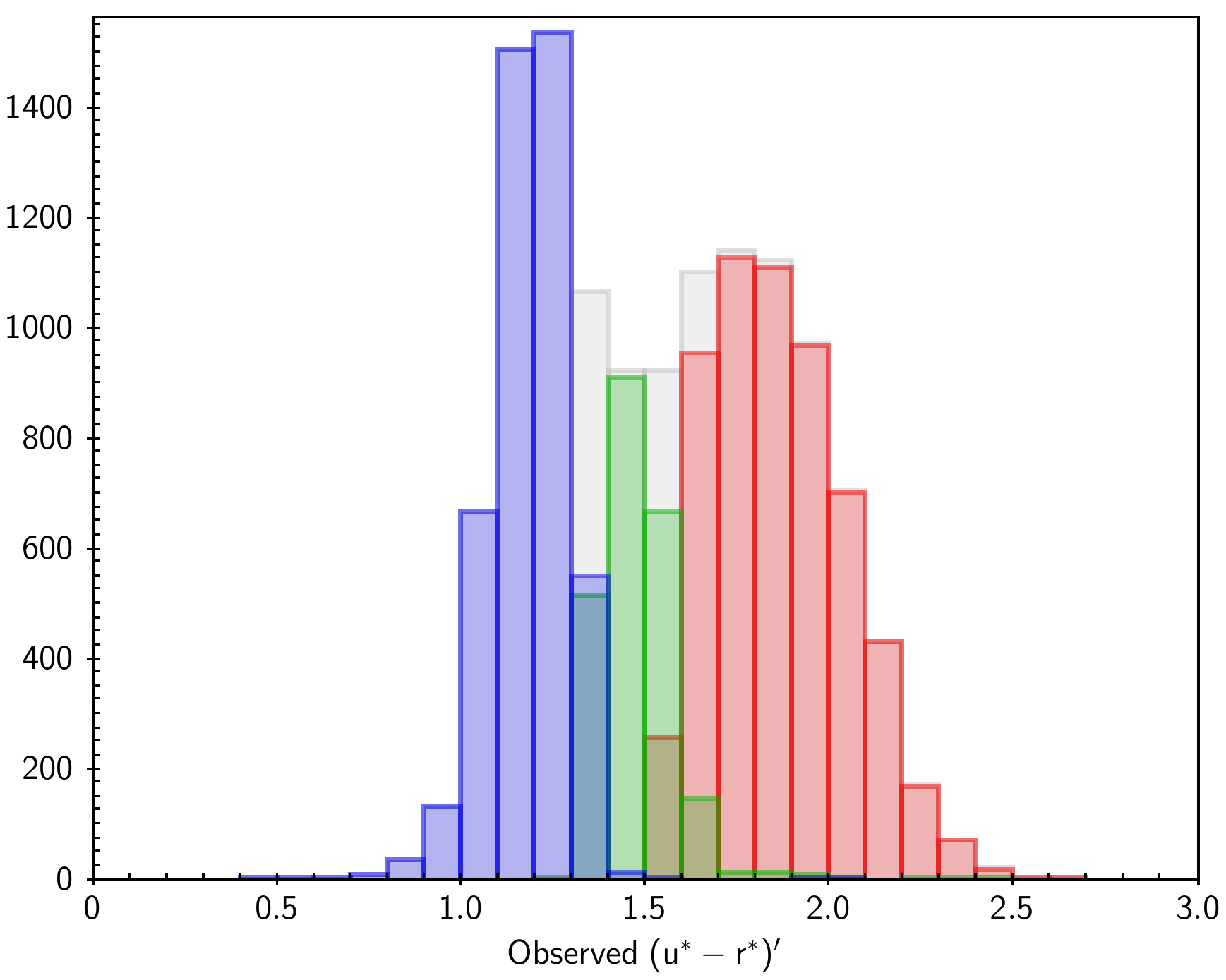}
\includegraphics[width=\linewidth]{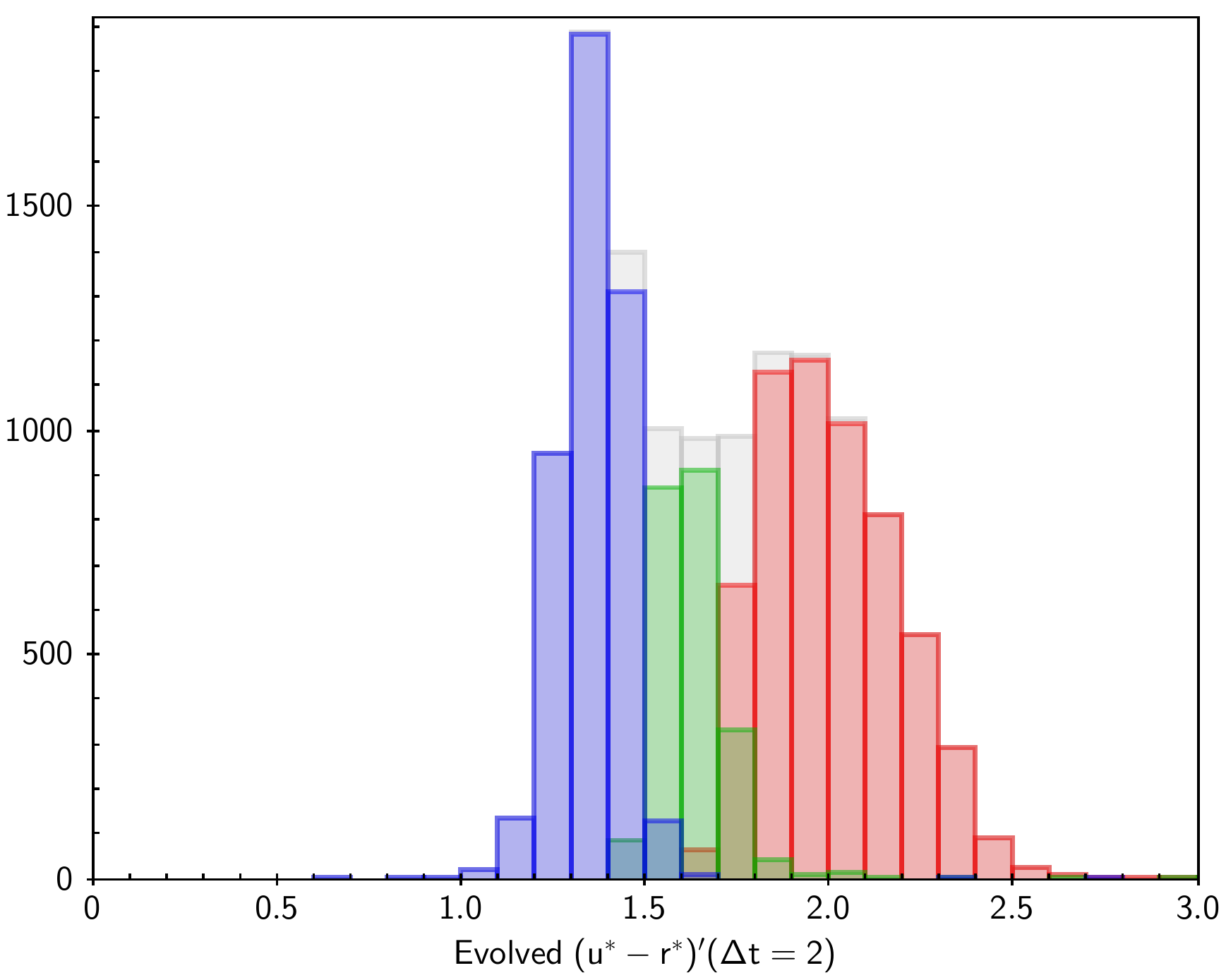}
\includegraphics[width=\linewidth]{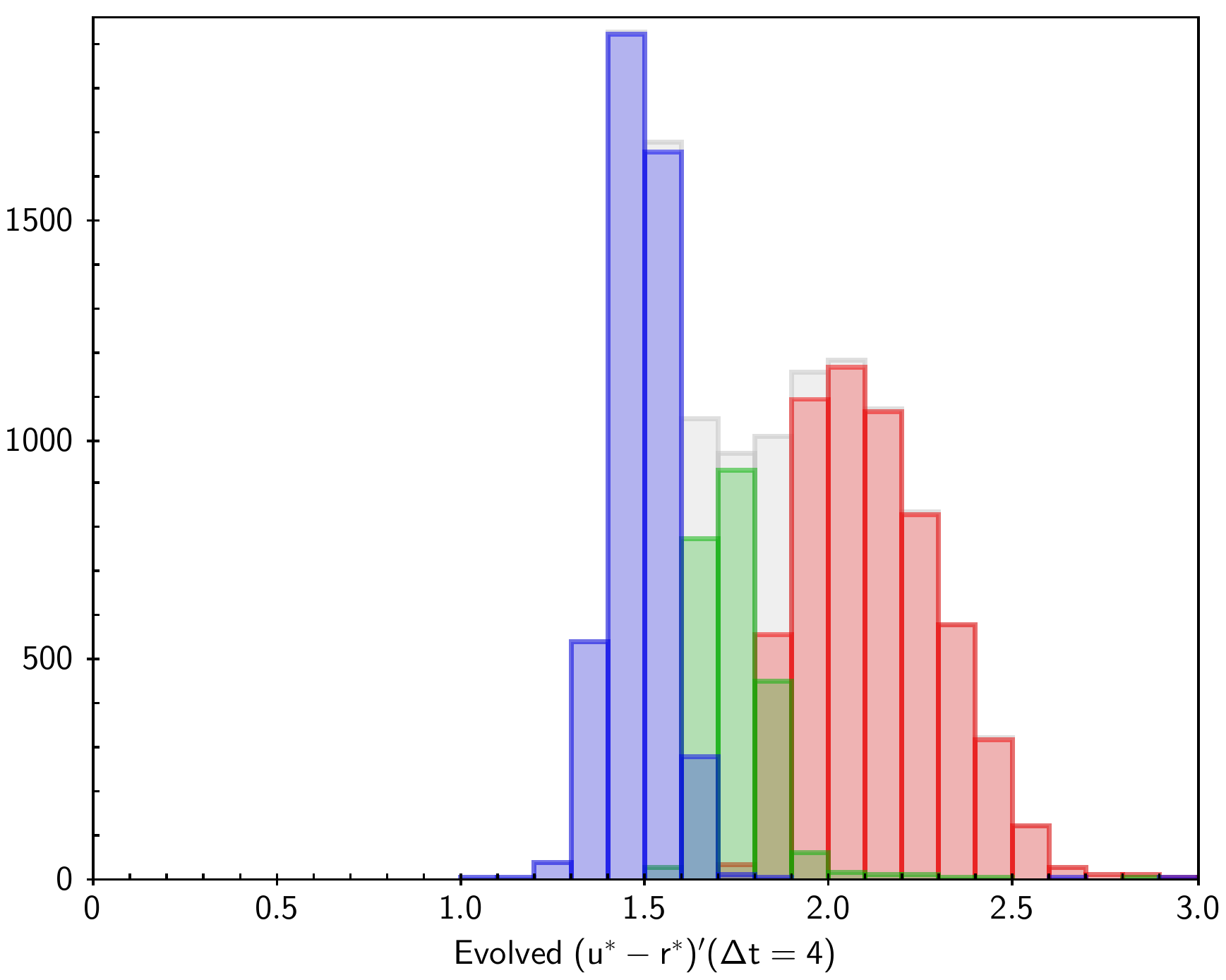}
\caption{Distributions of pseudocolour as observed (top) and evolved forward for $\Delta t = 2$ and $\Delta t = 4$ (middle and bottom) for the mass limited sample.}
\label{time_mass10_figures5}
\end{figure}

These timescales also apply for the mass-limited sample from Paper I. The most important difference for this subsample compared to our main one is the lack of very blue galaxies and particularly the corresponding decrease in the fraction of galaxies with very short $t/\tau$ (see Fig. \ref{time_t_tau_mass}). However this has essentially no effect on the evolution across the green valley. We still remove of order half the current green galaxies in $\sim 2$~Gyr and essentially all of them in $\sim 4$~Gyr (Fig. \ref{time_mass10_figures5}), just as for the complete sample.

Of course, all these estimates are somewhat dependent on the choice of green valley galaxies. If we had chosen narrower colour limits, then clearly galaxies would be able to transit the green colour range in a shorter time. In addition, we can see that the transit times are mass dependent. From the middle and bottom panels of Fig. \ref{time_synthcolour_mass_evolution}, we can see that the lowest mass green galaxies have just crossed the (diagonal) top edge of the current valley after 2~Gyr, while the highest mass objects only reach it after 4~Gyr. This is determined not only by the mass dependent SFH timescales (Figs. 9 and 10), but by the wider range of `green' colours at higher mass in our original selection (since the upper and lower colour limits are not parallel).

Recall that we have used a fixed slope for the evolution of the colour with log($t/\tau$) for all our galaxies, so strictly we have overestimated the shift of the colours for the galaxies with the longest and shortest $t/\tau$ values (bluest and reddest galaxies) since these evolve more slowly in colour, as noted earlier. Thus the top of the red distribution and the bottom of the blue distribution should really be moved downwards slightly in the two lower panels of  Fig. \ref{time_synthcolour_mass_evolution}, but this has no effect on the green galaxies (present or near future), which are well within the linear regime of our empirical colour correction in Fig. 11.

We should also reiterate that all these results are modulo a fixed $\tau$ for each galaxy; if evolution speeds up, then obviously these states can be reached after a shorter time, though not by a large factor if we assume that the $\tau$ for the current `just red' galaxies ($\tau \sim 3$~Gyr) places a lower limit on $\tau$ for green galaxies during their later evolution. Even if we were to assume that (somehow) $\tau$ in the future suddenly becomes 3~Gyr for all green valley galaxies this would clearly only scale down our calculated times by a factor $\sim 0.75$. (In fact, as discussed in the next section, there is no evidence for a recent `speeding up' of the evolution in green galaxies in our data). In either case, we can conservatively conclude that to evolve half to all of the current green galaxies out of the current valley will take 1.5 to 4~Gyr.

The timescales derived in this section differ by a factor $\sim 1.5$ from the earlier estimate of the green valley crossing time based on `typical' sSFRs of green and red galaxies, primarily because of the wide spread of sSFR at a given colour and the more precise consideration of the green valley boundaries. Since it is the change in colour that we are specifically interested in here, re. the definition of `green valley', the lower value of the timescale found from the galaxy by galaxy calculation should provide the better estimate. 

\subsection{A Low $z$ Sample}

As a sanity check, we can explore briefly the changes in colour across the redshift range surveyed by GAMA. To do this we choose an additional sample from the redshift range $0.02 < z < 0.06$ \citep[cf.][]{Moffett2016}, but with all other constraints the same as in our main sample.  In order that mass is not a complicating factor, we choose only galaxies with $10^{10} < M_{*}/M_{\odot} < 10^{11}$. (Using the full set of data available at low $z$ would include disproportionally large numbers of low luminosity blue galaxies compared to our main sample). Notice that this is a wider mass range than for our previous mass limited sample (and that in the low $z$ green valley study of Paper II), but the narrower range would leave too few low $z$ galaxies for a statistical study. With the wider limits we have 805 objects. To obtain the longest possible time difference, we then choose a sample in the same mass range and  at $0.19<z<0.20$ from our main sample. This contains 2996 objects.

\begin{figure}
\includegraphics[width=\linewidth]{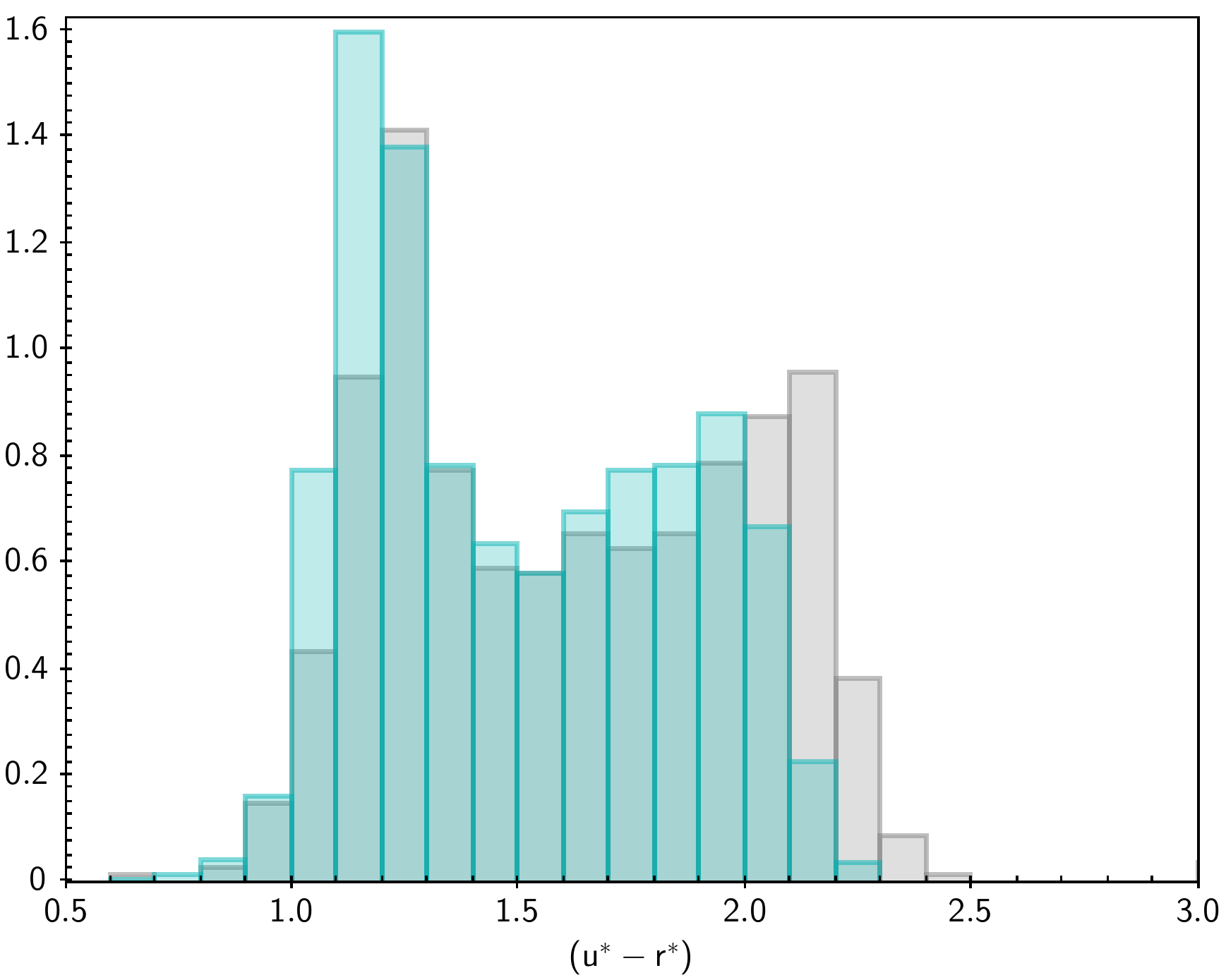}
\caption{Normalised distributions of observed $(u^* - r^*)$ colour for subsets of galaxies at $0.19<z<0.20$ (light blue) and $0.02<z<0.06$ (grey), as described in Section 3.5.}
\label{time_lowz_colour_hist}
\end{figure}

The look-back time difference between $z=0.02$ and $z=0.2$ is just over 2~Gyr, so would expect to see comparable differences in the samples to those illustrated in, e.g., Fig. 13. In this case we can use the actual (as opposed to pseudo-) colours, and we find that indeed, the overall distribution is shifted redwards by around 0.15 magnitudes in the low $z$ sample, as shown in Fig. \ref{time_lowz_colour_hist}, consistent with the predictions of our modelling (compare the top and middle panels of Fig. 13). The values of $t/\tau$ for the galaxies in the new valley are also increased by $\sim 0.5$, as expected if $\tau$ has remained typically 4~Gyr for these objects.  {\it Inter alia} this supports the physical reality of the derived parameters which we use.

Indeed, the close match of the distribution of the predicted evolved colours of the $0.19 < z < 0.20$ sample to the actually observed colour distribution at $0.02 < z < 0.06$ is strong empirical evidence for evolution with fixed $\tau$ in each galaxy, i.e. subsequent evolution of the $z=0.2$ galaxies following the same fall-off as in their preceding SFH. If ($z=0.2$) green galaxies suddenly accelerated their evolution in the last 2~Gyr we would expect to see more red galaxies, and even fewer in the valley, than is actually the case at low $z$. 

\begin{figure}
\includegraphics[width=\linewidth]{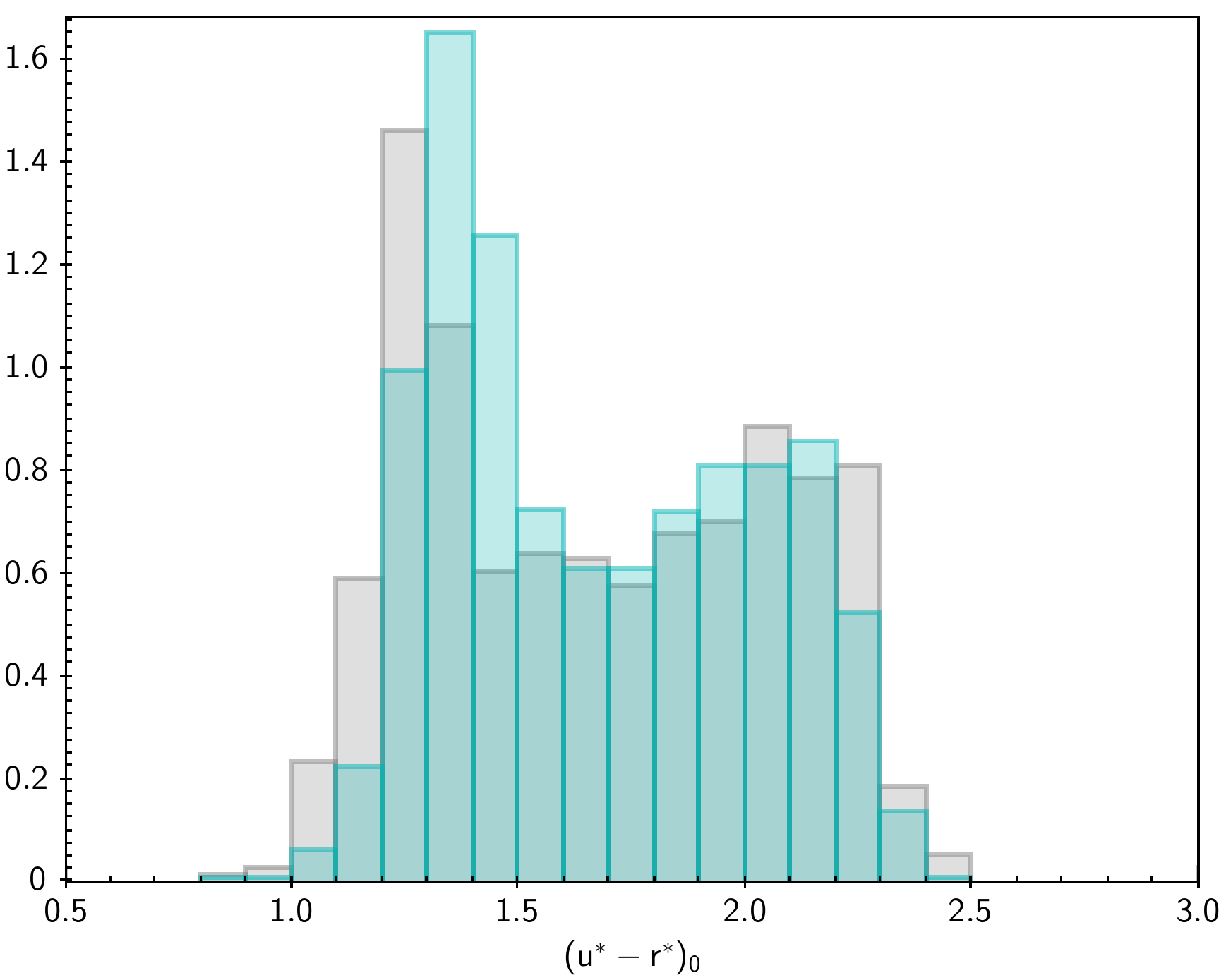}
\caption{Normalised distributions of $(u^* - r^*)$ colour for the subsets of galaxies at $0.19<z<0.20$ (light blue) and $0.02<z<0.06$ (grey) evolved to $z=0$ via their individual look-back times.}
\label{time_colour_zero_hist}
\end{figure}

Fig. \ref{time_colour_zero_hist} illustrates this further. Given that the low $z$ sample do not all correspond to a single epoch, we have here evolved the observed colours of the galaxies in both the $0.02 < z < 0.06$ and the $0.19 < z < 0.20$ samples to redshift zero via the look-back times to their individual redshifts. The agreement is very good (except for the bluest galaxies, for which we know we have overestimated the colour change, effectively moving the blue peak one bin too far to the right in the plot), giving further evidence that evolution of the the colours with a fixed $\tau$ for each galaxy gives a consistent result. A greater amount of evolution for the $z \simeq 0.2$ green galaxies, because of an accelerated decline in SF over the last 2~Gyr, would generate colours which would tend to be too red and, as noted previously, produce too large a dip at green colours. 

\subsection{How Green is the Valley?}

The predicted distribution of what we should see at later times also suggests that we ought to consider whether to allow for the fact that `a' green valley, as defined as a dip (or plateau) in the colour distribution, may move redwards as the overall sSFR distribution declines. Does the same range of colour or $t/\tau$ represent a valley at all epochs? Eventually, presumably, galaxies may reach the current green valley colour while still being amongst the bluest objects.

Looking for a dip in the pseudo-colour distributions, or at least the edge of the blue peak, we can see from Fig. \ref{time_synthcolour_evolution} that at $\Delta t = 2$~Gyr this will be at roughly $(u^* - r^*)' \simeq 1.75$ (compared to $\simeq 1.55$ currently), and by $\Delta t = 4$~Gyr it is at $(u^* - r^*) \simeq 1.85$. As noted above, very red and very blue galaxies do not follow the same $t/\tau$ versus colour trend that we have utilised in our evolution, since the colours pile up at the ends of the colour distribution regardless of the detailed SFH \citep[cf.][]{Eales2017, Eales2018, Oemler2017}. Allowing for this does not change our conclusions regarding the evolution of green galaxies in the next few Gyr, though a pile up at the reddest colours would emphasise the dip at green colours. 

By $\Delta t = 4$~Gyr, $t/\tau$ for typical current green galaxies will be 1.6 times its current value (as usual, assuming no change in $\tau$), also corresponding to $(u^* - r^*)' \simeq 1.7$. In that sense the current green galaxies are somewhat `marooned' in an apparent intermediate state as all three current colour subsets will evolve to redder colours. Unless they change their $\tau$, current green galaxies will always have sSFR in between those of the current red galaxies with little star formation and short $\tau$ and current blue galaxies with significant star formation and long $\tau$, even though they will have moved out of the {\em current} green valley. 

As they do evolve more than the reds in {\em colour}, they will {\em eventually} close up on the existing red galaxies, once their SFR becomes negligible, and essentially merge into the red sequence, all of whose galaxies will be asymptoting to late time old stellar population colours \citep{Eales2017}. How long this takes will evidently depend on how closely similar in colour we require galaxies on the red sequence to be. 

Reversing this and looking back in time, the same arguments would suggest that the valley between strongly and weakly star forming galaxies should have been at bluer colours. This is in agreement with, for instance, the data from \cite{Bell2004} who show that in COMBO-17 the colours of the blue peak, adjacent to the valley, and of red sequence galaxies both move redwards from $z \simeq 1$ towards the present day \citep[see also][]{Belli2018}. We will explore running the SFH models backwards and comparing to higher redshift data samples in a separate publication.

\section{Discussion, Conclusions and Caveats}
The primary outcomes of the above analysis are that current green valley galaxies largely share a common SFH and have reached a specific phase of their evolution in terms of the fall-off in their SFR. 

In terms of timescales, we find that (current) green galaxies can leave/cross the (current) green valley on timescales of 1.5-4~Gyr. This is similar to, but somewhat longer than the 1-2~Gyr timescale proposed, on the basis of the fraction of the population residing in the green valley, in Paper I, and comparable or slightly longer than the 2.6~Gyr suggested by \cite{Rowlands2018} from an analysis of spectroscopic indices of candidate transition objects. 

However, interestingly as we have no specific quenching events, it is very similar to or shorter than typical estimates of the timescale for the fading which follows the cutting-off of gas supplies (`strangulation' or `starvation') of star-forming galaxies, generally between 4 and 8~Gyr \citep{Bekki2002,Peng2015,Wheeler2014,Guo2017,Trussler2018}. The larger of these estimates are, rather than the green valley crossing time, more comparable numerically to the e-folding times of the SFR that are measured here for green (blue) galaxies of $\tau \sim 4$ (6)~Gyr, or indeed the ages (by our definition) of green galaxies, $t \sim 7$~Gyr. Indeed, it should be noted, that different authors mean different things by `quenching time', for instance the time to reach the green valley or reach the red sequence, the time since SF started to decline from some earlier constant level, or an e-folding time of the SF (within various SFH models). Clearly, each may result in quantitatively different times being deduced.

We should note, too, that sytematic errors in our SED fitting process and changes in the Bayesian priors may affect the absolute values we have quoted. However all {\it relative} comparisons, between colour subsamples for instance, remain true even if we substantially change our priors, and a strong correlation between colour and $t/\tau$ is robustly obtained. 

As noted earlier, in a simple closed box (presumably appropriate after infall is prevented, for instance) with an exponentially declining SFR, $\tau$ is equal to the time for the remaining fuel to be used up at the current rate \cite[e.g.][]{Hopkins2008}. Clearly direct measurements of the gas masses for our galaxies would be very useful to test the overall picture \cite[cf.][for the GASS survey]{Catinella2012}. Upcoming surveys such as DINGO \citep{Duffy2012} will provide this data. Given that the typical $t/\tau$ implies that current green valley galaxies have reduced their SFR by a factor around 5 since its peak, then in a simple Kennicutt-Schmidt type model, we would also expect the gas supply to have decreased by a similar factor.

On a similar timescale to galaxies leaving the valley, we might expect current blue galaxies to repopulate it. However, it is apparent that most blue galaxies have longer $\tau$ than green valley galaxies so would not replicate the (current) green galaxy evolutionary parameters even after fading, unless only those blue galaxies with $\tau$ at the short end of their range evolve to be green. However, due to the increased $t$ they {\em can} have similar $t/\tau$ to current green galaxies.

We should stress that the timescales calculated throughout this paper depend on the assumed exponentially declining model SFH with a fixed $\tau$ for each galaxy. However, we can note that even changing the typical $\tau$ of green galaxies to that observed in red galaxies does not grossly change the results (giving values at the low end of the range we quote), and that furthermore there is no evidence in the comparison with the low $z$ colour distribution for any systematic speeding up of the evolution of the green galaxies over time.

We should also remember that the present work largely refers to galaxies in the field and in low mass groups. There are no really large clusters in the GAMA areas. The timescales are therefore strictly only relevant to such environments. Within these, splitting the galaxies into grouped and isolated \citep{Robotham2011} we see virtually no differences in the $\tau$ and $t/\tau$ distributions for the green galaxies. Of course, it remains possible that galaxies in high density environments (which we do not sample here) evolve through their green phase more rapidly. We will explore this question elsewhere.

In Paper I it was shown that the majority of green galaxies (in the mass limited sample) had underlying structural parameters consistent with those of red galaxies. In particular, fitting the radial light profiles at red or near infrared wavelengths (effectively the mass surface density) by a single S\'{e}rsic (1968) component, the distributions of the index $n$ were strongly overlapping for the two subsamples. Blue galaxies on the other hand had significantly different structural parameters. It was therefore argued that most green galaxies could evolve directly onto the red sequence with no (further) change to their mass distributions. In particular, this was taken to support the view \citep[e.g.][]{Cheung2012, Fang2013, Taylor2015} that a significant spheroidal component needs to be in place {\em before} a galaxy becomes green.

\begin{figure}
\includegraphics[width=\linewidth]{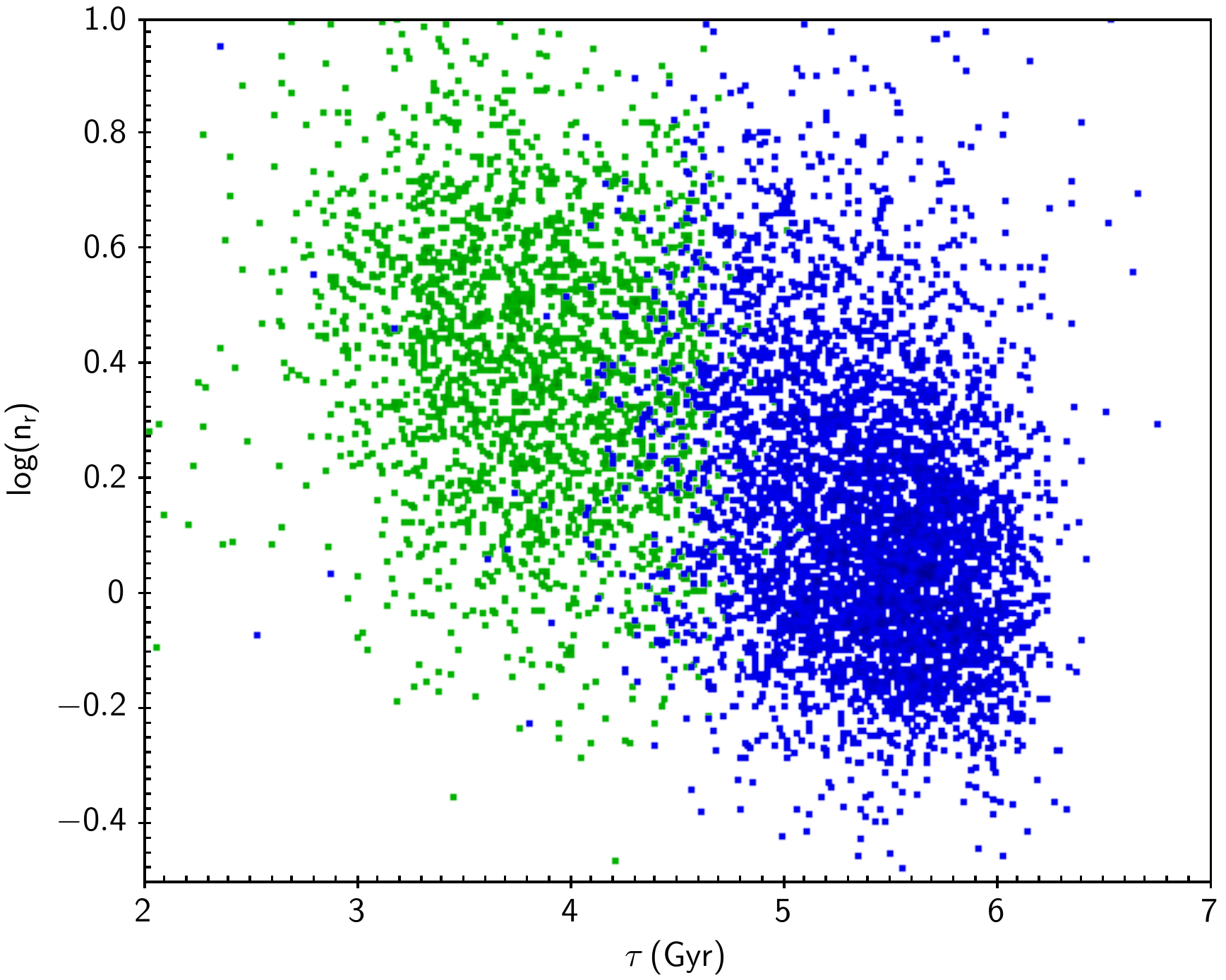}
\includegraphics[width=\linewidth]{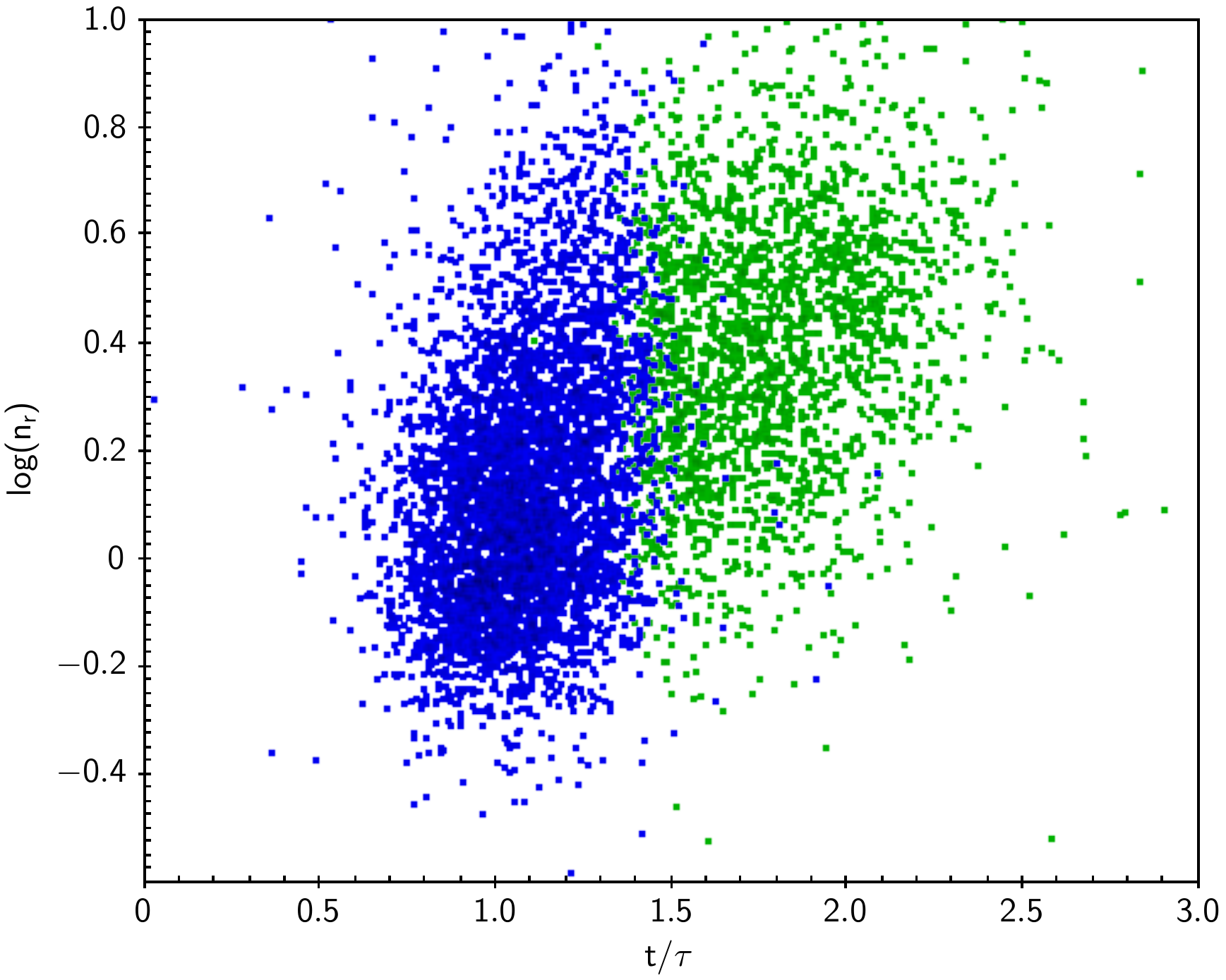}
\caption{Distributions of $\tau$ and $t/\tau$ against Sersic index for blue and green galaxies in the mass limited sample.}
\label{time_sersic_mass10}
\end{figure}

We can look for further evidence for such a scenario in the present data by looking at the distributions of S\'{e}rsic index $n$ against the stellar evolutionary parameters. We take the catalogued values of $n_r$ from \cite{Kelvin2012}. Fig. \ref{time_sersic_mass10} shows this plot for the mass-limited sample, so as to minimise any other influences (and because the mass range was originally chosen so as to well sample both blue and green galaxies). It is clear that the majority of the blue population has both long $\tau$ and small $n$, i.e. as expected they are disks with strongly ongoing SF. However, the blue galaxies with large $n$, say above 2.5 (log~$n > 0.4$), have on average shorter $\tau$ than the blue pure disks, but this is not by a large factor and is not enough to make them significantly overlap with the green population. Nevertheless, looking at the overall distribution for the combined blue and green populations does show a trend for shorter $\tau$ (and larger $t/\tau$) as $n_r$ increases, which is compatible with the effects of an increased bulge component changing both $n$ and the overall SFH of a two component system \citep[cf.][]{Lopez2018}. 

This is supported in a general way by the results from the smaller low $z$ sample, for which there exist morphological type classifications \citep[][Paper II]{Moffett2016}. Intermediate values of $\tau$, around 4~Gyr, typical of our main green sample, are seen primarily in the S0-Sa and E classes, i.e. in those galaxies with a large bulge componen, rather than in the Sab-Scd class. This agrees with the CALIFA results of \cite{Lopez2018} who find that S0/Sa galaxies have intermediate SFH timescales, and of \cite{Bitsakis2018} who show that CALIFA galaxies between the star forming main sequence and the quiescent region have morphological classifications from elliptical to early spiral. \cite{Chilingarian2017} likewise find that S0 and Sa galaxies have typical colours which place them in their green valley region.

The final caveat is that, depending how you look at it, `green' galaxies may never leave their valley. Since red galaxies pile up at their asymptotic colours at large ages and (at least currently) the blue galaxies decrease their star formation more slowly than other types, the present green galaxies may always remain as an intermediate population as the whole galaxy population reddens over time, at least until the point where they become completely red and dead, lose the imprint of their individual SFHs and merge into the red sequence.

\section*{Acknowledgements}

GAMA is a joint European-Australasian project based around a spectroscopic campaign using the Anglo-Australian Telescope. The GAMA input catalogue is based on data taken from the Sloan Digital Sky Survey and the UKIRT Infrared Deep Sky Survey. Complementary imaging of the GAMA regions is being obtained by a number of independent survey programs including {\it GALEX} MIS, VST KiDS, VISTA VIKING, {\it WISE}, {\it Herschel}-ATLAS, GMRT and ASKAP providing UV to radio coverage. 
 GAMA is funded by the STFC (UK), the ARC (Australia), the AAO, and the participating institutions. The GAMA website is http://www.gama-survey.org/. This work made extensive use of TOPCAT \citep{Taylor2005} software packages, which are supported by an STFC grant to the University of Bristol.

\section*{Appendix}

\begin{figure}
\includegraphics[width=\linewidth]{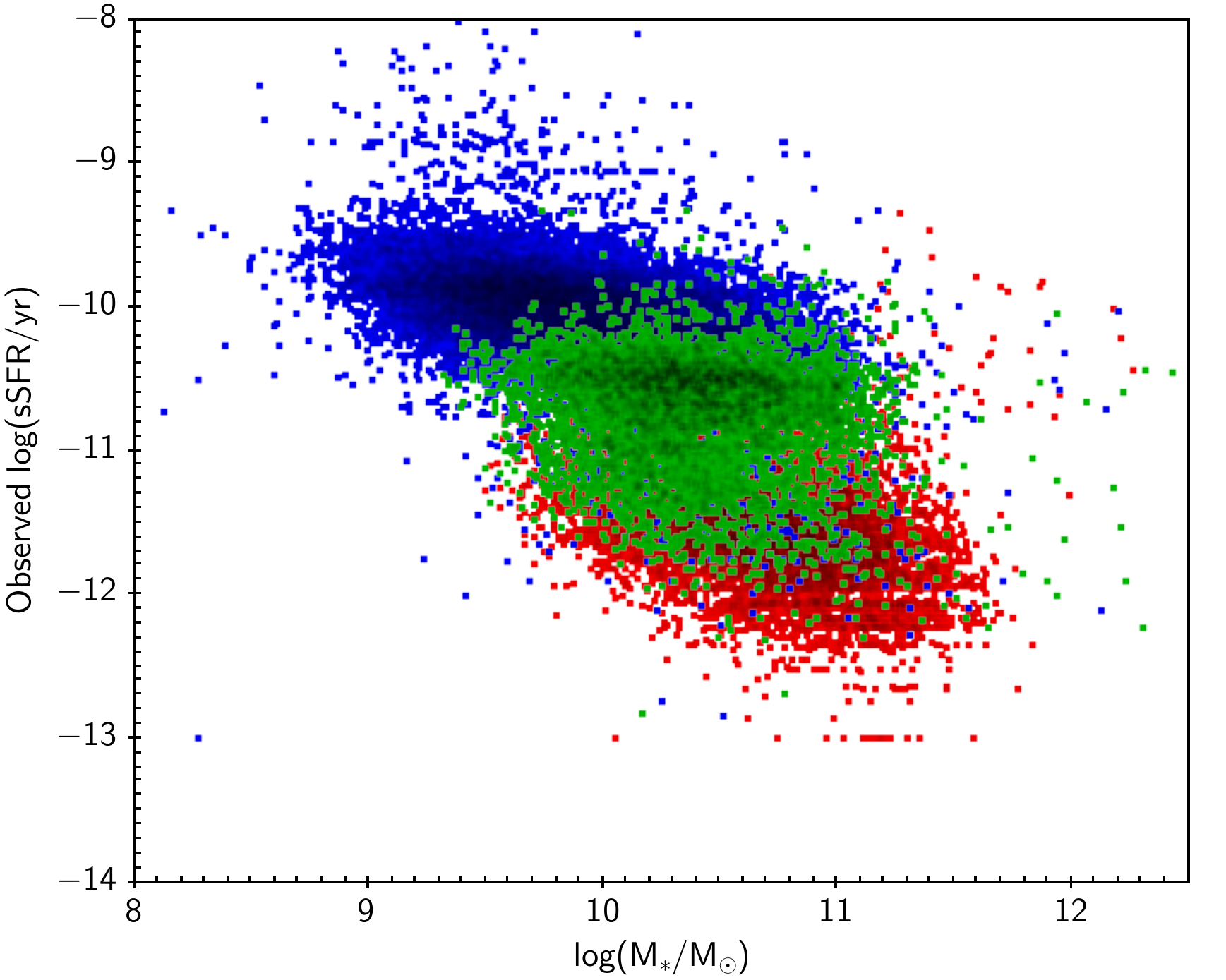}
\includegraphics[width=\linewidth]{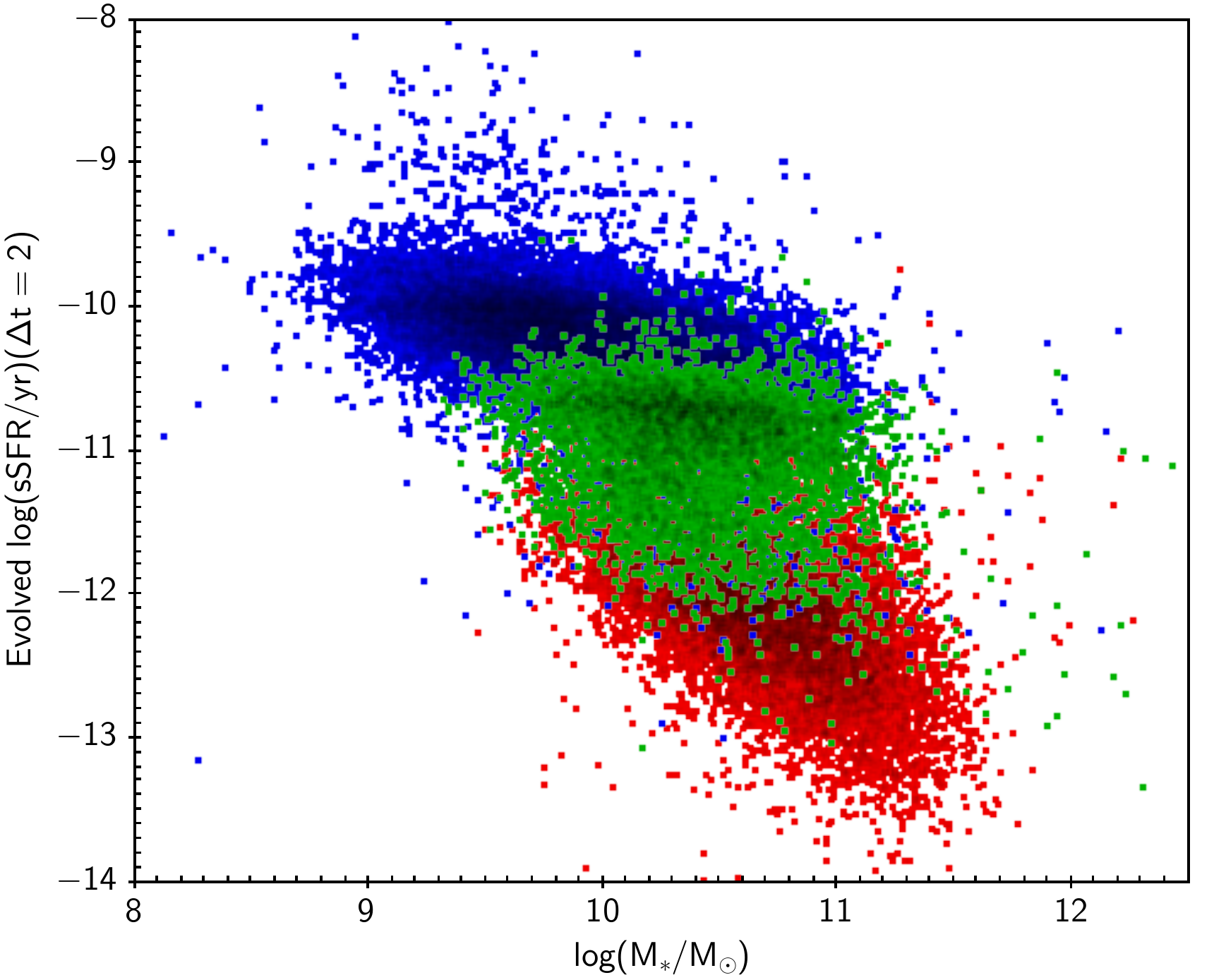}
\includegraphics[width=\linewidth]{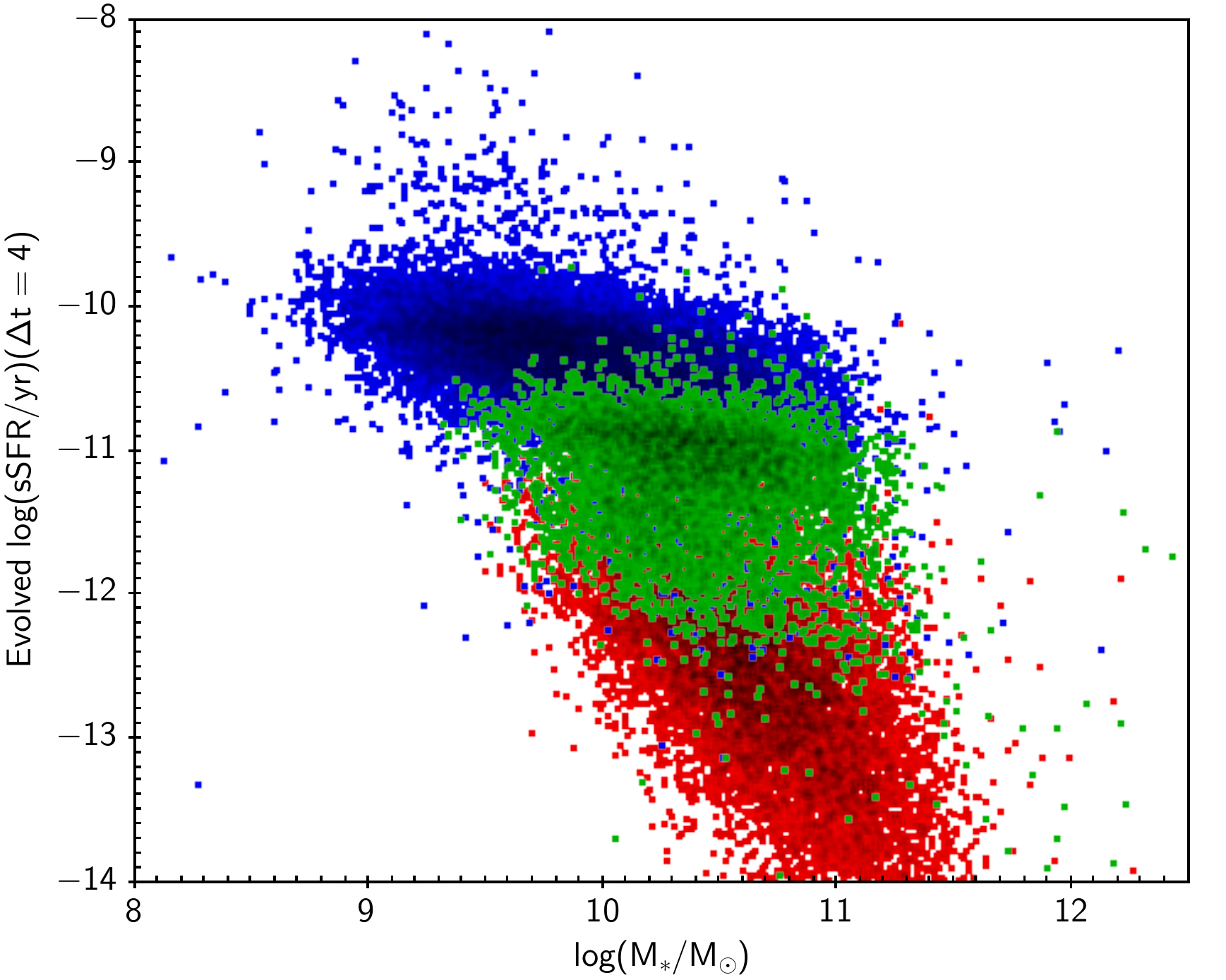}
\caption{Distribution of sSFR as a function of (observed) mass, (top) as observed (cf. Fig. 4), (middle) as evolved forwards by 2~Gyr and (bottom) as evolved forwards by 4~Gyr assuming an exponential fall-off with constant $\tau$ for each galaxy (see text).}
\label{time_ssfr_mass_evolution}
\end{figure}

For illustration, Fig. \ref{time_ssfr_mass_evolution} shows the observed and evolved sSFR distributions (Section 3.4) against observed stellar mass, i.e. not allowing for any increase in the stellar mass with time. This latter will only be significant for the bluest galaxies with nearly constant star formation rates as a function of time (long $\tau$) which could change their stellar masses by a few tenths of a dex over a 4~Gyr time span.

As expected, over time the current green galaxies decrease their typical sSFRs 
to $\sim 10^{-11}$yr$^{-1}$, with an extension down to $\sim 10^{-12}$, the majority continuing to lie at the edge of the blue cloud in SFR terms.

\label{lastpage}

\end{document}